\documentclass{article}

\usepackage{indentfirst}
\usepackage{starfont,amsfonts,amssymb}
\usepackage[reqno,intlimits]{amsmath}
\usepackage[polish,english]{babel}
\usepackage{bbold,scalerel}
\usepackage{microtype}
\usepackage{geometry}
\usepackage{graphicx}
\usepackage{fouriernc}
\DeclareMathAlphabet{\mathscr}{OMS}{zplm}{m}{n}

\newcommand{\dt}[1]{\overset{\raisebox{-0.1ex}[0ex][0ex]{\scaleobj{1}{.}}}{#1}}
\newcommand{\ddt}[1]{\overset{\raisebox{-0.1ex}[0ex][0ex]{.\hspace{-0.15ex}.}}{#1}}
\newcommand{\dddt}[1]{\overset{\raisebox{-0.1ex}[0ex][0ex]{.\hspace{-0.17ex}.\hspace{-0.17ex}.}}{#1}}

\makeatletter
\renewcommand{\@fnsymbol}[1]{\ensuremath{
    \ifcase#1\or \text{\Pentagram} \or \odot \else\@ctrerr\fi}}
\makeatother

\begin{document}

\title{Tensor $f(R)$ theory of gravity}

\author{
Tomasz Stachowiak\footnote{tomasz@amp.i.kyoto-u.ac.jp}\\
\small Department of Applied Mathematics and Physics,\\
\small Graduate School of Informatics, Kyoto University,\\
\small 606-8501 Kyoto, Japan}

\maketitle

\begin{abstract}
I propose an alternative $f(R)$ theory of gravity constructed by applying
the function $f$ directly to the Ricci tensor instead of the Ricci scalar. The
main goal of this study is to derive the resulting modified Einstein equations
for the metric case with Levi-Civita connection, as well as for the general
nonmetric connection with torsion. The modification is then applied
to the Robertson-Walker metric so that the cosmological evolution corresponding
to the standard model can be studied. An appealing feature is that even in the
vacuum case, scenarios without initial singularity and exponential expansion
can be recovered.  Finally, formulae for possible observational tests are given.
\end{abstract}

\section{Introduction}

The foundation of the present work is to consider a modified Lagrangian
(density), which depends functionally on the full Ricci tensor $R_{ab}$, not
just on its trace $\mathscr{R}$ as is the case in the so-called
$f(\mathscr{R})$ theories of gravity. The principles of relativity require
that this modification be obtained covariantly, and not component-wise, so
writing $f(R_{ab})$ could be misleading. Since $f$ will be a
tensor-valued function, for the sake of distinction from the usual
$f(\mathscr{R})$ theory, the extension will be referred to as tensor $f(R)$.

The motivation in both cases is the same -- the inclusion of
higher-order-of-curvature effects which can classically be ignored, but which
lead to
important modifications in other regimes. Most notably, the Starobinsky
inflation model \cite{Starobinsky} induced by
quadratic terms is a particularly important result in this spirit. Although
initially introduced on quantum gravity grounds, with corrections built from
various contractions of the Ricci tensor, it is now often considered in the
language of quadratic $f(\mathscr{R})$ theories \cite{Allemandi}.

Despite the initial similarity, the tensor $f(R)$ gravity presented here
differs considerably from the usual one, and the goal of this article is to
focus first on the development of this new theory, with a comparative study left for
future work. Accordingly, the notation and mathematical setting will be given
as well as the modified Einstein equations. Not to stop at the abstract level I
will also consider possible applications to cosmology, with a view to
nonsingular evolution, and provide basic formulae to be used in observational
cosmology.

Notable differences and similarities with the ordinary $f(\mathscr{R})$ theory will be
pointed out throughout the derivations in Sections \ref{sect_2}, \ref{Palatini},
and \ref{sec_FRW}, but for a more complete, general review of the standard
approach, the reader might want to consult review articles \cite{Olmo},
\cite{Nojiri}, or \cite{Faraoni} and references therein.

\section{Construction of the modified action}
\label{sect_2}

In the usual $f(\mathscr{R})$ theories one postulates the Lagrangian
\begin{equation}
    \mathcal{L}_0=f(\mathrm{tr}[R]) = f\left(R_{ac} g^{ca}\right),
\end{equation}
with the summation convention used, and the covariant metric tensor denoted by
$g_{ac}$. On purely abstract grounds, the order in which $f$ and trace
appear is not fixed, so instead of the above I will consider the Lagrangian
to be
\begin{equation}
    \mathcal{L}_g=\mathrm{tr}[f(R)] = {[f(R)]}_{ac} g^{ca},
\end{equation}
where the square brackets are used to indicate elements of a matrix, and the
bare symbol $R$ has to refer to the tensor not the scalar, as explained below.

A similar idea has been studied before by Borowiec
\cite{Borowiec1999,Borowiec2003}, but it differed from the present work in two
ways. First, it used a torsionless metric and second, the Lagrangian depended
on polynomial
invariants of the Ricci tensor $\text{tr}[R^k]$. Such scalars formed
with powers $k$ higher than the space-time dimension can be reduced to the
lower ones by using the characteristic polynomial. However, this cannot, in
general, be done explicitly for transcendental functions -- i.e., when one needs
to use an infinite series of powers of $R$. What is more, the coefficients of
the characteristic polynomial themselves depend on the components of $R$, leading
to an unwieldy expression of an original function of $R$ in terms of a function
of the invariants $\text{tr}[R^k]$. The present work aims at overcoming this
problem, and also at including connections with the most general torsion and
nonmetricity.
 
To proceed with the general treatment, the first thing to settle is what
tensors and operators to use, and in particular how to interpret $f(R)$. Power
series immediately come to mind, so what is needed is a representation of $R$
such that it can be composed with itself by matrix multiplication consistent
with relativistic index contraction. In other words, $R$ should be an
endomorphism, for instance on the tangent bundle over the space-time.

To treat $R$ as such an endomorphism, mixed indices
have to be used so that the result composition ${R^a}_b{R^b}_c$ is again a
mixed-indices tensor of the same valence. ${R_a}^b$ would do as well, but with the
former choice the eigenvalue problem can be written as
\begin{equation}
    {R^a}_b v^b = \lambda v^a,
\end{equation}
i.e., for eigenvectors rather than eigenforms, which seems more natural. The
two are still equivalent through the musical isomorphism, and such an $R$ 
is a self-adjoint operator with regard to the metric
\begin{equation}
    \langle u,R(v)\rangle = u^a g_{ab} {R^b}_c v^c = u^a R_{ca}v^c
    = {R^b}_a u^a g_{bc}v^c = \langle R(u),v\rangle,
\end{equation}
provided that $R_{ab}$ is symmetric, which is the case for the Levi-Civita
connection. When one allows for the torsion to be nonzero the above requires a
generalization given in Section \ref{Palatini}.

In the bracket-component notation, $f$ should act on $R$ considered as a linear
operator with matrix elements ${[R]^a}_b$, and should also give as the result an
operator, whose elements are denoted by ${[f(R)]^a}_b$. For example, for the
composition with itself it is convenient to write 
${[R\cdot R]^a}_b = {[R^2]^a}_b$, so the superscript 2 refers to the operator power, not a
component. Accordingly, $R$ will signify the $(1,1)$ valence tensor, and for the
Ricci scalar the contraction ${R^a}_a$ or $\mathscr{R}$ will be used. After
``bracketing,'' the index notation is recovered, which allows for raising and
lowering; for brevity, the brackets will be omitted in the simplest cases such
as $[R]_{ab}=R_{ab}$.

For any analytic $f:\mathbb{R}\rightarrow\mathbb{R}$ the following
definition of the matrix function 
$f^*:\mathbb{R}^{n^2}\rightarrow\mathbb{R}^{n^2}$ can be \mbox{used
\footnote{As with $R$, it will be convenient to sometimes write the identity
operator without indices, so in order to avoid confusion with the variation
$\delta$, I will use ${\mathbb{1}^a}_c$ instead of the Kronecker symbol
${\delta^a}_c$.}:}
\begin{equation}
\begin{aligned}
    {[f^*(R)]^a}_b &:= \sum_{n=0}^{\infty} f_n {[R^n]^a}_b\\
        &= f_0 {\mathbb{1}^a}_b + f_1 {R^a}_b+f_2 {R^a}_s {R^s}_b + 
        f_3 {R^a}_s {R^s}_t {R^t}_b + \cdots,
\end{aligned}
\end{equation}
where
\begin{equation}
    f(\xi) = \sum_{n=0}^{\infty}f_n \xi^n, \label{ser}
\end{equation}
and the sums are written explicitly, as they are not tensor contractions ($R^n$
is an operator power as explained above). The above requires that the spectral radius of 
$\rho(R) = \lim\limits_{n\rightarrow\infty} \|R^n\|^{1/n}$ be
less than the radius of convergence of the series $f(\xi)$.

For example, when $f=\exp$, the above two Lagrangians are
\begin{equation}
\begin{aligned}
    \mathcal{L}_0 &= \exp\left(\mathscr{R}\right) 
    = 1+ \mathscr{R}+\frac{1}{2!}\mathscr{R}^2
    +\frac{1}{3!}\mathscr{R}^3+\cdots\\
    &= 1 + {R^a}_a + \frac{1}{2!}({R^a}_a)^2 
    +\frac{1}{3!}({R^a}_a)^3+\cdots,\\
    \mathcal{L}_g &= \mathrm{tr}\left[\mathbb{1} + R + \frac{1}{2!}R^2 + 
    \frac{1}{3!}R^3 +\cdots\right]\\
    &= d + {R^a}_a + \frac{1}{2!}{R^a}_b{R^b}_a + 
    \frac{1}{3!}{R^a}_b{R^b}_c{R^c}_a + \cdots,
\end{aligned}
\end{equation}
where $d$ is the dimension of the space-time. Thus the first essential
deviation appears at the quadratic level and is proportional to
$f_2({R^a}_b{R^b}_a-({R^a}_a)^2)$ if the same $f$ is used in both approaches.
The difference is also evident
when the Lagrangians are written in terms of the eigenvalues of $R$:
\begin{equation}
    \mathcal{L}_0 = f\left(\sum_i \lambda_i\right)\quad\text{vs}\quad
    \mathcal{L}_g = \sum_i f(\lambda_i).
    \label{eigen_act}
\end{equation}
A degeneracy in $\lambda_i$ might then lead to the same theories, e.g., when the
traceless Ricci tensor vanishes:
$\hat{R}^a_{\phantom{a}b}:={R^a}_b-\frac{1}{d}\mathscr{R}{\mathbb{1}^a}_b=0$.
The Ricci tensor is then
proportional to the identity matrix and ${[f(R)]^a}_a=d f(\mathscr{R}/d)$,
 which, up to a simple rescaling of $f$, is the same as the Lagrangian
$\mathcal{L}_0$.
However, one has to be careful when making such substitutions directly in
the action, because $R$ is determined only after having solved the Einstein
equations. If the assumption $\hat{R}=0$ is justified from the beginning, the
two theories coincide. We shall see in the examples below that even in an empty
universe this condition might not hold generally but just for isolated solutions. 

Note also -- that if $f$ is determined, there is no freedom of choice for its
constant term $f_0$, which naturally corresponds to the cosmological constant.
In other words, in such a nonperturbative interpretation, its value is tied to the
whole expansion and cannot be adjusted independently. The expansion around
$R=0$ also shows that when $f$ is almost linear, then higher-order terms can be
ignored in the weak field limit leading to the Einstein-Hilbert action and a
small perturbation of general relativity.

Although intuitive, the above definition is not very convenient when a function
is real analytic but has complex singularities like $\tanh(\xi)$. A
definition better suited for the situation at hand is an elegant generalization
of Cauchy's formula\footnote{In what follows, $f$ and $f^*$ can safely be treated
as the same object, so the star will be dropped.}
\begin{equation}
    f(R) := \frac{1}{2\pi\mathrm{i}}\int_C 
    (\xi\mathbb{1} -R)^{-1}f(\xi)\;\mathrm{d}\xi,
\end{equation}
for a contour $C$ which encloses the spectrum of $R$ but not the singularities
of $f(\xi)$. The two definitions agree for fairly general assumptions, and for a
function that is real on the real axis, the matrix $f(R)$ will also be real
\cite{Higham}. 

The dimension of $f(R)$ affects how the function is given, because $R$ has the
units of curvature, and so should the Lagrangian. At first, it seems two
constants are necessary to give $f(R) = C_0 \tilde{f}(R/C_1)$ in terms of a
function $\tilde{f}$ which only contains dimensionless parameters, but this can
be rewritten as
\begin{equation}
    f(R) = C_0 \tilde{f}\left(\frac{C_0}{C_1}\frac{R}{C_0}\right)
    \;\rightarrow\; C_0\tilde{f}(R/C_0),
\end{equation}
with a redefined dimensionless $\tilde{f}$. The remaining constant $C_0$ can
then be further rescaled using the cosmological or the Hubble constant
depending on context -- this is done in Section \ref{sec_FRW}.

Having defined $\text{tr}[f(R)]$, the total action, including the matter
Lagrangian $\mathcal{L}_M$, is taken to be 
\begin{equation}
    \mathscr{S} = \int\left( \frac{1}{16\pi\mathscr{G}}\mathcal{L}_g +
    \mathcal{L}_M\right) \sqrt{-g}\mathrm{d}^4x,
\end{equation}
where $\mathscr{G}$ is the gravitational constant, and the modified Einstein
equations can then be obtained in one of the two standard
ways. One is to assume the Levi-Civita connection and take the metric as the
dynamical variable; the other is to consider both the metric and the connection
as dynamical. The former is called the metric and the latter the Palatini
formulation (or, more generally, metric-affine).

In both cases the variation of the $f(R)$ term is needed, and the second
definition of a tensor function allows us to easily calculate it as
\begin{equation}
\begin{aligned}
    \delta\mathrm{tr}[f(R)] &= 
    \text{tr}\left[\frac{1}{2\pi\mathrm{i}}\int_C
    (\xi\mathbb{1}-R)^{-1}\delta R(\xi\mathbb{1}-R)^{-1}f(\xi)\;\mathrm{d}\xi\right]
    = \text{tr}\left[\frac{1}{2\pi\mathrm{i}}\int_C
    (\xi\mathbb{1}-R)^{-2}f(\xi)\delta R\;\mathrm{d}\xi\right]\\
    &= \text{tr}\left[\frac{1}{2\pi\mathrm{i}}\int_C
    (\xi\mathbb{1}-R)^{-1}f'(\xi)\;\mathrm{d}\xi\;\;\delta R\right]
    = \text{tr}\left[f'(R)\delta R\right],
\end{aligned}
\end{equation}
where the cyclic property of trace
\begin{equation}
    \mathrm{tr}[X_1 X_2\ldots X_k] = \mathrm{tr}[X_k X_1 X_2\ldots X_{k-1}],
\end{equation}
was used in the first line, and integration by parts in the second.
Reexpressing $\delta R$ with $\delta g$ and $\delta\Gamma$ to arrive at
the modified Einstein equation is the subject of the next two sections.

\subsection{Definitions and notation}
To shortly review the conventions used, the covariant derivative and the
connection coefficients in a basis $\{e_a\}$ are related through
\begin{equation}
    \nabla_{e_a}e_b = {\Gamma^c}_{b a}e_c,
\end{equation}
so that for a coordinate basis $e_a = \partial_a$ one has
\begin{equation}
\nabla_a X^b = \partial_a X^b + {\Gamma^b}_{ca}X^c.
\end{equation}
As $\Gamma$ will not in general be symmetric in the lower indices, care needs
to be taken regarding their order. The antisymmetric part of the connection
defines the torsion as
\begin{equation}
    T(X,Y) := \nabla_X Y - \nabla_Y X - [X,Y] = e_a {T^a}_{bc}X^bY^c,
\end{equation}
and in a coordinate basis, where $[\partial_a,\partial_b]=0$, it follows that
\begin{equation}
    {T^a}_{bc} = {\Gamma^a}_{cb} - {\Gamma^a}_{bc}.
\end{equation}

The Riemann tensor is given by 
\footnote{The brackets involving vectors denote commutation not
antisymmetrization -- i.e.,
there is no prefactor of $\frac12$.}

\begin{equation}
    \mathsf{R}(X,Y)Z := \nabla_{[X}\nabla_{Y]} Z -
    \nabla_{[X,Y]}Z
    = e_d{\mathsf{R}^d}_{abc}Z^aX^bY^c,
\end{equation}
or, in term of components in a coordinate basis,
\begin{equation}
    {\mathsf{R}^d}_{abc} = \partial_b{\Gamma^d}_{ac}-\partial_c{\Gamma^d}_{ab}
    +{\Gamma^d}_{sb}{\Gamma^s}_{ac}-{\Gamma^d}_{sc}{\Gamma^s}_{ab},
\end{equation}
and the Ricci tensor is the contraction
\begin{equation}
    R_{ab} := {\mathsf{R}^c}_{acb}.
\end{equation}
Note, then that although $R_{ab}$ is constructed solely with the connection
(curvature), for the operator ${R^a}_b = g^{ac}R_{cb}$ the metric is necessary.
Finally, the signature will be taken to be $(-,+,+,+)$, and the speed of light
equal to unity, so that coordinates have the dimension of length, and the
metric itself is dimensionless.

\section{The metric approach}

The natural connection solely determined by the metric through $\nabla_a
g_{bc}=0$ and ${T^a}_{bc}=0$ is the Levi-Civita connection. Its variation, 
as expressed by $\delta g$, is 
\begin{equation}
    \delta{\Gamma^c}_{ba} = \frac12 g^{cd}(\nabla_b\delta g_{ad}+
    \nabla_a\delta g_{db} - \nabla_d\delta g_{ba}),
\end{equation}
and in turn for the covariant Ricci tensor one has
\begin{equation}
    \delta R_{ab} =
    \nabla_c(\delta{\Gamma^c}_{ab})-\nabla_b(\delta{\Gamma^c}_{ac}),
\end{equation}
which accordingly gives
\begin{equation}
\begin{aligned}
    \delta R_{ab} &= \frac12 \left(\nabla^d\nabla_b\delta g_{ad}+
    \nabla^d\nabla_a\delta g_{db} - \nabla^d\nabla_d\delta g_{ba}\right)+\\
    &\phantom{=} -\frac12\left(g^{cd}\nabla_b\nabla_a\delta g_{dc}
    +\nabla_b\nabla^d\delta g_{ad}
    -\nabla_b\nabla_d g^{cd}\delta g_{ca} \right)\\
    &= \frac12\left(\nabla^d\nabla_b\delta g_{ad} + \nabla^d\nabla_a\delta g_{bd}
    - \Box\delta g_{ab} - g^{cd}\nabla_b\nabla_a\delta g_{cd}\right).
\end{aligned}
    \label{dRab}
\end{equation}

Next, by observing that
\begin{equation}
    0 = \delta({\mathbb{1}^a}_c) = g_{bc}\delta g^{ab} + g^{ab}\delta g_{bc},
\end{equation}
the variation of the operator $R$ becomes
\begin{equation}
    \delta{R^a}_b = g^{ac}(\delta R_{cb} -{R^s}_b\delta g_{cs}),
\end{equation}
leading to
\begin{equation}
\begin{aligned}
    \delta\left(\text{tr}[f(R)]\sqrt{-g}\right) 
    &= \left({[f'(R)]^a}_b\delta{R^b}_a 
    +\tfrac12\text{tr}[f(R)]g^{bd}\delta g_{bd}\right)\sqrt{-g}\\
    &= \left([f'(R)]^{ac}\delta R_{ca} -[R f'(R)]^{cd}\delta g_{dc}
    +\tfrac12\text{tr}[f(R)]g^{bd}\delta g_{bd}\right)\sqrt{-g}.
\end{aligned}
\end{equation}

The variation $\delta R_{ab}$ of \eqref{dRab} can be substituted
into the above, and due to 
$\sqrt{-g}\;\nabla_a X^a = \partial_a\left(\sqrt{-g}X^a\right)$
each term containing the covariant derivative can be integrated by parts
provided that the variations vanish at the boundary or that the boundary is
empty. The result is
\begin{equation}
\begin{aligned}
    \delta\left(\text{tr}[f(R)]\sqrt{-g}\right) &=
    \left(\nabla_c\nabla^d[f'(R)]^{cb} - \tfrac12 \Box [f'(R)]^{bd} - 
    \tfrac12 \nabla_a\nabla_c [f'(R)]^{ac}g^{bd}\right.\\
    &\phantom{=}\left. - [Rf'(R)]^{bd}
    +\tfrac12 {[f(R)]^a}_a g^{bd}\right)\sqrt{-g}\delta g_{bd}.
\end{aligned}
\end{equation}

Finally, defining the stress-energy tensor $\mathcal{T}$ by
\begin{equation}
    \frac{\delta(\sqrt{-g}\mathcal{L}_M)}{\delta g_{bd}}
    =: \frac12 \mathcal{T}^{bd}\sqrt{-g},
\end{equation}
the condition $\delta\mathscr{S}=0$ gives the following modified Einstein
equations 
\begin{equation}
\boxed{
\begin{aligned}
    \tfrac12\Box [f'(R)]_{bd}-\nabla_c\nabla_b{[f'(R)]^c}_d
    +\tfrac12\nabla^a\nabla^c[f'(R)]_{ac}g_{bd}&\phantom{=}\\
    +[Rf'(R)]_{bd}-\tfrac12\text{tr}[f(R)]g_{bd}
    &= 8\pi\mathscr{G}\mathcal{T}_{bd}.
\end{aligned}}
\label{metric_ein}
\end{equation}
As can be seen, the last two terms on the left-hand side reduce to the standard
Einstein tensor for $f=\mathrm{Id}$, whereas the other terms are zero since
$f'=1$.

\section{The Palatini approach}\label{Palatini}

In the more general case, the connection is independent of the metric, and there
are two assumptions that can be relaxed here: vanishing torsion and 
metric compatibility. In general the connection can be decomposed into the sum
\begin{equation}
    {\Gamma^a}_{bc} = \widetilde{\Gamma}^a_{\phantom{a}bc} + {K^a}_{bc}
    -{C^a}_{bc},\qquad
    K_{abc} := -\frac12\left(T_{abc}+T_{bca}-T_{cab}\right),
    \label{gen_con}
\end{equation}
where $\widetilde\Gamma$ is the Levi-Civita connection for $g$, $K$ is
called the contorsion tensor, and $C$ describes the nonmetricity
\begin{equation}
    C_{abc} := \frac12\left(\nabla_cg_{ab}+\nabla_bg_{ca}
    -\nabla_ag_{bc}\right).
\end{equation}

Accordingly, the variation of the Ricci tensor is now
\begin{equation}
    \delta R_{ab} =
    \nabla_c(\delta{\Gamma^c}_{ab})-\nabla_b(\delta{\Gamma^c}_{ac})
    -{T^d}_{bc}\delta{\Gamma^c}_{ad},
\end{equation}
and neither the connection coefficients nor the Ricci tensor are symmetric in
the lower indices. The eigenvalues of $R$ might not be real any more, in which
case they appear in conjugate pairs. This means that the
trace of $f(R)$ will still be real, for real analytic $f$.

There is, however, a possible natural generalization, because of the following
identity\footnote{The underline denotes the sum over cyclic permutations.}
\begin{equation}
    R_{ab} = R_{ba} + \nabla_{\underline{a}} {T^c}_{\underline{cb}} 
    + {T^c}_{cd} {T^d}_{ab},
\end{equation}
which leads to the introduction of a new tensor, which is the symmetric part of
$R$,
\begin{equation}
    S_{ab} := R_{ab} -\frac12(\nabla_{\underline{a}} {T^c}_{\underline{cb}}
    + {T^c}_{cd}{T^d}_{ab}).
\end{equation}
These tensors have the same trace so there is no need for $S_{ab}$ in the
standard $f(\mathscr{R})$ theories -- the trace cancels the imaginary parts of the
conjugate pairs of the eigenvalues. Here, the situation is different, because
the function $f$ is applied to the eigenvalues of $R$ before the trace is
taken, so although the final result is real, it also depends on the imaginary
parts. The other reasons and equations for the $f(S)$ variant are given
following the $f(R)$ derivation below.

In contrast to the preceding section, only first derivatives are present in the
action, and the integration by parts requires an additional term, because the
torsion affects the expression for covariant divergence:
\begin{equation}
    \sqrt{-g}\;\nabla_a\left(X^a\right) = \partial_a\left(\sqrt{-g}\;X^a\right) 
    + \sqrt{-g}\left({T^b}_{ba} - {C^b}_{ba}\right)X^a.
\end{equation}
The total variation of the Lagrangian then becomes
\begin{equation}
\begin{aligned}
    \delta\left(\text{tr}[f(R)]\sqrt{-g}\right) &= 
    \left(P^{ba}\delta R_{ab} -[R f'(R)]^{db}\delta g_{bd}
    +\tfrac12{[f(R)]^a}_a g^{bd}\delta g_{bd}\right)\sqrt{-g}\\
    &= \left(\tfrac12\text{tr}[f(R)]g^{bd}-[Rf'(R)]^{db}\right)
    \delta g_{bd}\sqrt{-g}\\
    &\phantom{=} +\left( \nabla_b P^{ba}{\mathbb{1}^d}_c
    -\nabla_cP^{da}
    -{T^d}_{bc}P^{ba}\right.\\
    &\phantom{=+}\left.
    +({C^s}_{sb}-{T^s}_{sb})P^{ba}{\mathbb{1}^d}_c
    -({C^s}_{sc}-{T^s}_{sc})P^{da}
    \right)\delta{\Gamma^c}_{ad}\sqrt{-g},
\end{aligned}
\label{derivation}
\end{equation}
where the derivative tensor is denoted by $P_{ab}:=[f'(R)]_{ab}$ for brevity.

In addition to the stress-energy tensor $\mathcal{T}$, a new quantity is
necessary to reflect the fact that matter fields can, in general, depend on the
connection -- if only through the covariant derivative. The hyper-momentum
tensor is defined thus:
\begin{equation}
    \sqrt{-g}\;{\mathscr{Q}_a}^{bc} := \frac{\delta\left(\sqrt{-g}\mathcal{L}_M\right)}
    {\delta{\Gamma^a}_{bc}},
\end{equation}
and the modified Einstein equations can now be written as
\begin{equation}
    \boxed{
\begin{aligned}
    8\pi\mathscr{G}\mathcal{T}_{bd} &= 
    \lbrack Rf'(R) \rbrack_{(db)} - \tfrac12 \text{tr}[f(R)]g_{bd},\\
     8\pi\mathscr{G}{\mathscr{Q}_c}^{ad} &= 
     \nabla_b\left({\mathbb{1}^{[b}}_cP^{d]a}\right)
      -{(T-C)^s}_{sb}{\mathbb{1}^{[b}}_cP^{d]a}
      -\tfrac12{T^d}_{cb}{P}^{ba},
\end{aligned}}
\end{equation}
where the symmetrization is necessary, because the variation $\delta g_{bd}$ is
symmetric, even though $R_{bd}$ is not.

The second set of equations can be simplified if an auxiliary connection is
defined to be
\begin{equation}
    \hat{\Gamma}^a_{\phantom{a}bc} := {\Gamma^a}_{bc} -
    \tfrac12 {\mathbb{1}^a}_{b}{(T-C)^s}_{sc},
\label{aux_con}
\end{equation}
and using the associated covariant derivative $\hat\nabla$, the second set of
Einstein equations reads
\begin{equation}
     8\pi\mathscr{G}{\mathscr{Q}_c}^{ad} = 
     \hat\nabla_b
     \left({\mathbb{1}^{[b}}_{c}P^{d]a}\right)
     -\tfrac12{T^d}_{cb}P^{ba}.
\end{equation}
Additionally, contraction over the pair of indices $\{cd\}$ leads to 
\begin{equation}
    3\hat\nabla_bP^{ba}
    +{T^s}_{sb}P^{ba}+16\pi\mathscr{G}{\mathscr{Q}_s}^{as} = 0,
\end{equation}
which allows us to rewrite the main equations as
\begin{equation}
\begin{aligned}
    \lbrack Rf'(R) \rbrack_{(db)} - \tfrac12 \text{tr}[f(R)]g_{bd}
    &= 8\pi\mathscr{G}\mathcal{T}_{bd},\\
    \hat\nabla_c P^{da}
    -\left({T^d}_{cb}-\tfrac13{T^s}_{sb}{\mathbb{1}^d}_c\right)P^{ba}
     &= 16\pi\mathscr{G}\left({\mathscr{Q}_c}^{ad}
     -\tfrac13{\mathscr{Q}_s}^{as}{\mathbb{1}^d}_c\right).
     \label{traced_eq}
\end{aligned}
\end{equation}

As in the ordinary $f(R)$ formulation, the torsion equations become algebraic
for the Einstein-Hilbert case $f(R)=R$ because $P^{ab}=[f'(R)]^{ab}=g^{ab}$, so
that derivatives of $\Gamma$ only appear in $R$. Further, if the matter fields
are such that $\mathscr{Q}_{abc}\equiv 0$, contractions of the torsion
equations give
\begin{equation}
\begin{aligned}
     3{C^s}_{sa} = -6{C_{as}}^s &= 2{T^s}_{sa},\\
     2C_{(ad)c} &= T_{dac} +\tfrac23 {T^s}_{s(a}g_{c)d}
\end{aligned}
\end{equation}
This means that if ${T^s}_{sa}=0$, then $2C_{(ad)c}=T_{dac}$, and it
follows immediately from \eqref{gen_con} that $K_{abc}=C_{abc}$. But that, by
definition, means the connection must be the Levi-Civita one.

In other words, for $f(R)=R$, zero hyper-momentum and totally antisymmetric
torsion, the theory becomes standard general relativity.
Note that for this to happen it is not necessary to assume zero torsion from
the beginning, just that all its traces vanish.

Since the Ricci tensor is, in general, no longer symmetric, the tensor $P$
cannot be used directly to define a new metric for which equation
\eqref{traced_eq} would define a metric connection. In the standard
$f(\mathscr{R})$ theories, the tensor that enters is $R$ itself, and it can be decomposed
into (anti)symmetric parts at the level of the Einstein equations, as the
function $f$ is applied only to its trace, and all $f({R^a}_a)$ terms are just
scalars.

Here, the situation is different in that even in the first set of equations the
symmetrization is applied to $Rf'(R)$, not to $R$, and the second
set of equations contains $f(R)$, not $f'(R)$. Because even for the second power
one has $g^{bc}X_{c(d}X_{a)b}\neq X_{(ab)}g^{bc}X_{(cd)}$, symmetrizing the
equations would not lead to a single distinguished tensor to be used as the new
metric. Moreover, even though
the components of $R$ are real, it seems natural to consider a self-adjoint
matrix, for which the action is directly related to the eigenvalues as in
\eqref{eigen_act}.

These problems could be overcome by constructing the action with the symmetric
tensor $S$, introduced before, whose variation is simply
$\delta S_{ab} = \frac12(\delta R_{ab} + \delta R_{ba})$. The derivation is
essentially the same as in \eqref{derivation}, and the difference is that the
tensor contracted with $\delta g$ is already symmetric, so the Einstein
equations are
\begin{equation}
\begin{aligned}
    8\pi\mathscr{G}\mathcal{T}_{bd} &= 
    \lbrack S f'(S) \rbrack_{db} - \tfrac12 \text{tr}[f(S)]g_{bd},\\
8\pi\mathscr{G}{\mathscr{Q}_c}^{ad} &= 
     \hat\nabla_b
     \left({\mathbb{1}^{[b}}_{c}P^{d]a}\right)
     -\tfrac12{T^d}_{cb}P^{ba},
\end{aligned}
\label{sym_ricci_ein}
\end{equation}
where now, by a slight abuse of notation, $P_{ab} = [f'(S)]_{ab}$, and the
auxiliary covariant derivative is the one given by equation \eqref{aux_con}.

As before, the trace can be used to rewrite the second equation as
\eqref{traced_eq}, and following the same reasoning as for the standard $f(R)$
derivation, the torsionless connection with no hyper-momentum yields
\begin{equation}
    \hat{\nabla}_c P^{da} = 0.
    \label{pseudometr}
\end{equation}
This would indicate that $\hat\Gamma$ is the Levi-Civita
connection for the metric $P^{da}$, but the situation is complicated by the
fact that the tensor $P=f'(R)$ is not conformally related to the original
metric $g$, so the signature might not be the same, and the determinant of $g$ is
not directly proportional to that of $P$; also, raising of indices in $P$
does not amount to matrix inversion. It should also be kept in mind that
with the standard extension of covariant derivative to tensor densities, which
uses $\sqrt{|g|}$ to cancel the weight, the above equation can be rewritten as
\begin{equation}
    \frac{1}{\sqrt{|g|}}\hat\nabla_c\left(\sqrt{|g|}\;P^{da}\right) = 0,
\end{equation}
but this is not equivalent to
\begin{equation}
    \frac{1}{\sqrt{|\det{P}|}}\hat\nabla_c
    \left(\sqrt{|\det{P}|}\;P^{da}\right) = 0,
\end{equation}
unless $\sqrt{\det{P}}$ is used to extend $\hat\nabla$ to densities. Without
specifying which extension is used, the condition
$\nabla_c(\sqrt{|g|}g_{ab})=0$ does not necessarily indicate metricity,
contrary to what can sometimes be found in the literature. Because
of this freedom, and given the problems with inverting
$P$, the more fundamental equation \eqref{pseudometr} is better
as an indication of a metric connection in the present case.

With some effort, the Christoffel formula can be used to express $\hat\Gamma$ as
a function of derivatives of $P$, but the derivatives of the
connection coefficients are still involved in the nonlinear term $f'(S)$. The
question is then whether they can be eliminated with the help of the remaining
equations.

In the standard approach, the first set of the Einstein equations
\eqref{sym_ricci_ein} can, in principle, be used to
solve for the Ricci scalar and accordingly simplify the second set by using
the Ricci tensor associated with the new metric and its Levi-Civita connection
\cite{Olmo}. Here, one would have to solve nonlinear equations for the whole
tensor $S$ in order to eliminate the connection in the same manner. At
present, it appears that this path of investigation is not applicable, because
the equations involve full tensors $R$ or $S$, not just their traces.

\section{FRW dynamics}
\label{sec_FRW}

The standard cosmological model is the basic example that needs to be
considered in order to gain insight into the applicability of the proposed
modification. The model assumes spatial homogeneity and isotropy, requiring the
Robertson-Walker geometry, which in spherical coordinates 
$\{t,r,\theta,\varphi\}$ has the metric
\begin{equation}
    \text{d}s^2 = -\text{d}t^2 + a(t)^2\left(\frac{\text{d}r^2}{1-kr^2}
    +r^2\text{d}\Omega^2\right)=g_{ab}\text{d}x^a\text{d}x^b,
\label{RW}
\end{equation}
where $\text{d}\Omega^2=\text{d}\theta^2+\sin^2\theta\text{d}\varphi$ is the
standard metric on the unit sphere. The final assumption in this first attempt
at modified cosmology will be that the RW metric provides the only dynamical
variable -- the scale factor $a(t)$ -- the connection is that of Levi-Civita
and the metric formalism can be used.

Accordingly,
the matter source will be taken to be a homogeneous perfect fluid with density
$\rho$ and pressure $p$, so that the stress energy tensor is
\begin{equation}
    \mathscr{T}_{ab} = p g_{ab} + (p+\rho) u_a u_b,
\end{equation}
where the four-velocity in these coordinates is just $u=\partial_t$.

There are then effectively only two modified Einstein equations, one of third
order and one of fourth corresponding to the $\mathscr{T}_{00}$ and
$\mathscr{T}_{11}$ components of \eqref{metric_ein} respectively. However, the
latter follows from the
derivative of the former, which is the generalization of the Friedmann
equation
\begin{equation}
    \boxed{
\begin{aligned}
    H\left(3f''(\lambda_0)+f''(\lambda_1)\right)
    \frac{\text{d}\lambda_0}{\text{d}t}
    &= 16\pi\mathscr{G}\rho
    +6H^2\left((\lambda_1-\lambda_0)f''(\lambda_1)
    +f'(\lambda_1)-f'(\lambda_0)\right)\\
    &\phantom{=} +\lambda_0\left(f'(\lambda_0)+f'(\lambda_1)\right)
    -f(\lambda_0)-3f(\lambda_1)\\
\end{aligned}}
\label{gen_Fred}
\end{equation}
where $\lambda$ are the eigenvalues of $R$
\begin{equation}
    \lambda_0 = 3\frac{\ddt{a}}{a},\qquad
    \lambda_1 = 2H^2 + \frac{2k}{a^2}+\frac{\ddt{a}}{a},
\end{equation}
$H$ is the Hubble ``constant'' $H=\dt{a}/a$, and the overdot denotes the time
derivative.

The present value of the constant, $H_0:=H(0)$, is customarily used to obtain
dimensionless quantities and, as discussed in Section \ref{sect_2}, there is
still an unspecified constant in the function $f$. Although $H_0^2$ has the
suitable dimension, it will not do as $C_0$, because the function $f$ should be
a fundamental quantity valid for all gravitational actions, not just the FRW
cosmology, and thus cannot be defined with such specific constants. Instead,
$C_0$ will become a physical parameter of the new theory, and the Hubble
constant $H_0$ will serve to provide the dimensionless counterpart
$c_0:= C_0 H_0^{-2}$.

Of course, the roles could be reversed, with $C_0$ used instead of $H_0$, but
for initial clarity it is
better to keep with the convention of rescaling densities, time, etc., with
$H_0$. The dimensionless eigenvalues are then
\begin{equation}
    \alpha:=\lambda_0 H_0^{-2},\qquad \beta:=\lambda_1 H_0^{-2},
\end{equation}
which gives e.g. $f(\lambda_0) = f\left(\alpha H_0^{-2}\right)$ and leads to
further simplification
\begin{equation}
    H_0^{-2}f(\lambda_0) = c_0
    \tilde{f}(\alpha/c_0) =: F(\alpha),
\end{equation}
and similarly for $\beta$. The main equation can then be rewritten as
\begin{equation}
\begin{aligned}
    h(3F''(\alpha)+F''(\beta))\frac{\text{d}\alpha}{\text{d}\tau}&=
    6\Omega+6h^2\left((\beta-\alpha)F''(\beta) + F'(\beta)-F'(\alpha)\right)\\
    &\phantom{=}+\alpha(F'(\alpha)+F'(\beta)) - F(\alpha)-3F(\beta),
\end{aligned}
\label{modfried}
\end{equation}
where $h$, the density parameter and dimensionless time are defined by
\begin{equation}
    h := \frac{H}{H_0},\quad
    \Omega := \frac{8\pi\mathscr{G}\rho}{3H_0^2}\quad
    \text{and}\quad  \tau := H_0 t.
\end{equation}

The function $F$ can then be specified with any suitable number of
dimensionless parameters including $c_0$. It could be considered to be given
{\it a priori} by some elementary function like $A\sin(B \xi)$, or defined by
infinitely
many expansion coefficients as the series \eqref{ser}. Yet to consider such
coefficients as independent parameters would be to multiply entities
beyond necessity, so I will adopt the former approach here.

A quantitative reason can also be given for this, in anticipation of the
observational analysis. Finding the coefficients from the data
would undoubtedly lead to better and better fits as the number of
coefficients increases, but such a fit would come with a huge cost as measured
by the Akaike or Bayesian information criteria, which are now standard tools
of observational cosmology \cite{Liddle,Kurek}.

As for the nature of parameters in the present case, some more information can
be gleaned from the zeroth- and first-order expansions of $F$, as they reproduce
the standard model with the cosmological constant. The general
form\footnote{As before, $\xi$ is just an auxiliary independent variable used to
define functions and their rescalings.} is
$F(\xi) = F_0 + F_1 \xi$, but the overall rescaling of the Lagrangian is not
important, and taking $F_1=1$ gives the ordinary Friedmann equation
\begin{equation}
    H^2 = \frac{8\pi\mathscr{G}\rho}{3} -\frac{k}{a^2} 
    - \frac23 H_0^2F_0,
\end{equation}
upon identifying the cosmological constant $\Lambda = -2H_0^2F_0$. In terms of
the original function $f$, this means that $f_0=-\Lambda/2$, and it suggests
that the cosmological constant itself could be used as a fundamental
dimensional quantity by
\begin{equation}
    f(\xi) = \frac{\Lambda}{2}\tilde{f}\left(\frac{2\xi}{\Lambda}\right),
\end{equation}
with $\tilde{f}$ carrying no other free parameters. Using the respective
density parameter $\Omega_{\Lambda}:=\Lambda/3$, this means that
\begin{equation}
    F(\alpha) = \frac{1}{H_0^2}f(\lambda_0) =
    \frac{3\Omega_{\Lambda}}{2}
    \tilde{f}\left(\frac{2\alpha}{3\Omega_{\Lambda}}\right),
    \label{oneparamF}
\end{equation}
where the expansion of $\tilde{f}$ is then necessarily restricted to
\begin{equation}
    \tilde{f}(\xi) = -1 + \xi +\mathscr{O}(\xi^2).
\end{equation}

Turning now to the dynamics of this model, a minimal set of variables yielding
a closed system can be built from the derivatives of $a(t)$, or rather their
rescaled versions $h$ and $\alpha$,
which are identically related by $\alpha=3(\dt{h} +h^2)$. Also, the other of
the eigenvalues can be eliminated through
\begin{equation}
\beta = 2h^2 +\frac{2\Omega_k}{a^2}+\frac{\alpha}{3},\quad \text{with}\quad
\Omega_k := \frac{k}{H_0^2},
\end{equation}
although
for shorter notation it will be better to keep the symbol $\beta$ and
understand it as a function of $a$, $h$ and $\alpha$, which will be the
replacements for $\dt{a}$, $\ddt{a}$ and $\dddt{a}$.

Because the conservation law $\nabla^a\mathscr{T}_{ab}=0$ still holds, the
matter-energy density $\rho$ is expressible in terms of $a$ if one assumes
an equation of state
\begin{equation}
    p = (\gamma-1) \rho \quad\Longrightarrow \quad \rho(t) 
    = \rho_0 a(t)^{-3\gamma} \quad\Longrightarrow\quad
    \Omega = \sum_j \Omega_j a^{-3\gamma_j}.
\end{equation}
Finally, introducing
\begin{equation}
    W := 6h^2\left((\beta-\alpha)F''(\beta) + F'(\beta)-F'(\alpha)\right)
    +\alpha(F'(\alpha)+F'(\beta)) - F(\alpha)-3F(\beta),
\end{equation}
for the sake of brevity, a dynamical system with three
degrees of freedom described by the variables $\{\alpha,h,a\}$ is obtained:
\begin{equation}
\left\{\begin{aligned}
    \dt{\alpha} 
    &= \frac{6\Omega +W(\alpha,\beta,h,a)}
    {\left(3F''(\alpha)+F''(\beta)\right)h} &=: v_1(\alpha,h,a),\\
        \dt{h} &= \tfrac13\alpha -h^2 &=: v_2(\alpha,h,a),\\
        \dt{a} &= a h &=: v_3(\alpha,h,a),
        \label{maineq}
\end{aligned}\right.
\end{equation}
where the dot now refers to the new time $\tau$. Note that the denominator of
$v_1$ would only be identically zero for the purely linear $F$, which is the
standard general relativity. The form of $v_2$ and $v_3$ is dictated by the
definition of $h$ and the essential dynamics lies with $v_1$. This is also
where we find the difference in complexity between the new theory and
$f(\mathscr{R})$, for which $v_2$ and $v_3$ are the same, but the first
equation would read
\begin{equation}
    \dt{\alpha} = 3(\beta-\alpha)h+
    \frac{6\Omega-F(3\beta+\alpha)+2\alpha F'(3\beta+\alpha)}
    {12hF''(3\beta+\alpha)}.
\end{equation}
The difference between the two equations in the simplest quadratic case $F(\xi)
= -\frac32\Omega_{\Lambda}+\xi +F_2\xi^2$ is just 
$(\Omega_{\Lambda}+\Omega-h^2)/(2F_2 h)$, which is nonzero exactly when
the evolution deviates from the Friedmann equation. As was mentioned in Section
\ref{sect_2}, if $\hat{R}$ vanishes, then a simple rescaling of $F$ also leads
to the same equations but in this particular geometry the condition is very
restrictive. For flat universes (as in the examples below) the only solutions with
this property are the de Sitter ones, $h=\text{const}$, which do not exhaust all
possible solutions, even when $\Omega=0$. On the other hand, the difference disappears
completely if we take different functions: $F(\xi) = F_1 + \xi +F_2\xi^2$ for
$f(R)$ and $\widetilde{F}(\xi) = 4F_1 + \xi + 3F_2\xi^2$ for $f(\mathscr{R})$;
the theories are equivalent for the Robertson-Walker geometry at the
quadratic level, even when $\hat{R}\neq0$. However, no such simple
relation could be found for cubic terms.

A general feature of the main system \eqref{maineq} is that if the geometry is
flat, i.e.,
$k=0$ and the density does not depend on the scale factor, like for the
cosmological constant, then the first two equations decouple and give a planar
system. In fact, one could simply assume that no ordinary matter enters the
equations as $\Omega$, but instead consider the higher-order
terms of $F$ as some sort of field imitating matter. For example, if
$F(\xi) = -\frac32\Omega_{f}+\xi+\tfrac12F_2\xi^2$, the main
equation \eqref{modfried} becomes
\begin{equation}
    h^2 = \Omega_{f} +F_2\left(\dt{h}^2-2h\ddt{h}-\tfrac{14}{3}h^2\dt{h}
    +\tfrac{16}{9}h^4\right),
\end{equation}
so that $\Omega_f$ acts as dark energy and the $F_2$ term acts as effective
material content.

Another general, and problematic, feature of the $\dt{\alpha}$ equation
is the singularity at $h=0$, i.e., when expansion changes to
contraction and vice versa. This is not a singularity of equation
\eqref{modfried} and can lead to a valid solution provided that the numerator
of $v_1$ vanishes as well. Thus, care has to be taken when using the 
dynamical system form, because the singularities might simply
signify that the left-hand side of the original equation is zero, and vice
versa: a zero of $v_1$ might in fact be a singularity of the original equation
\eqref{maineq}.

\subsection{Examples of cosmological models}

\begin{figure}[h]
\includegraphics[width=.5\textwidth]{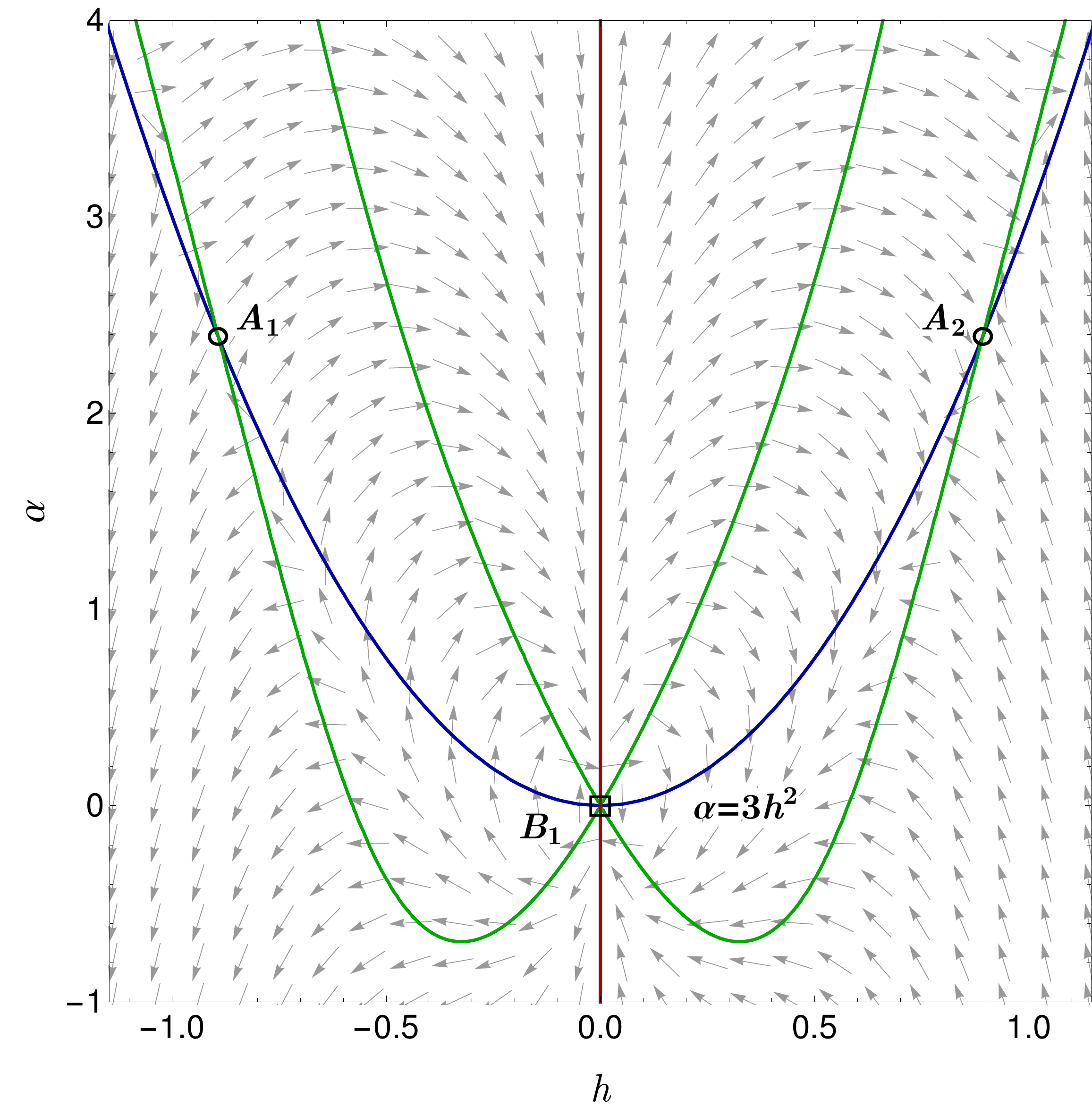}
\includegraphics[width=.5\textwidth]{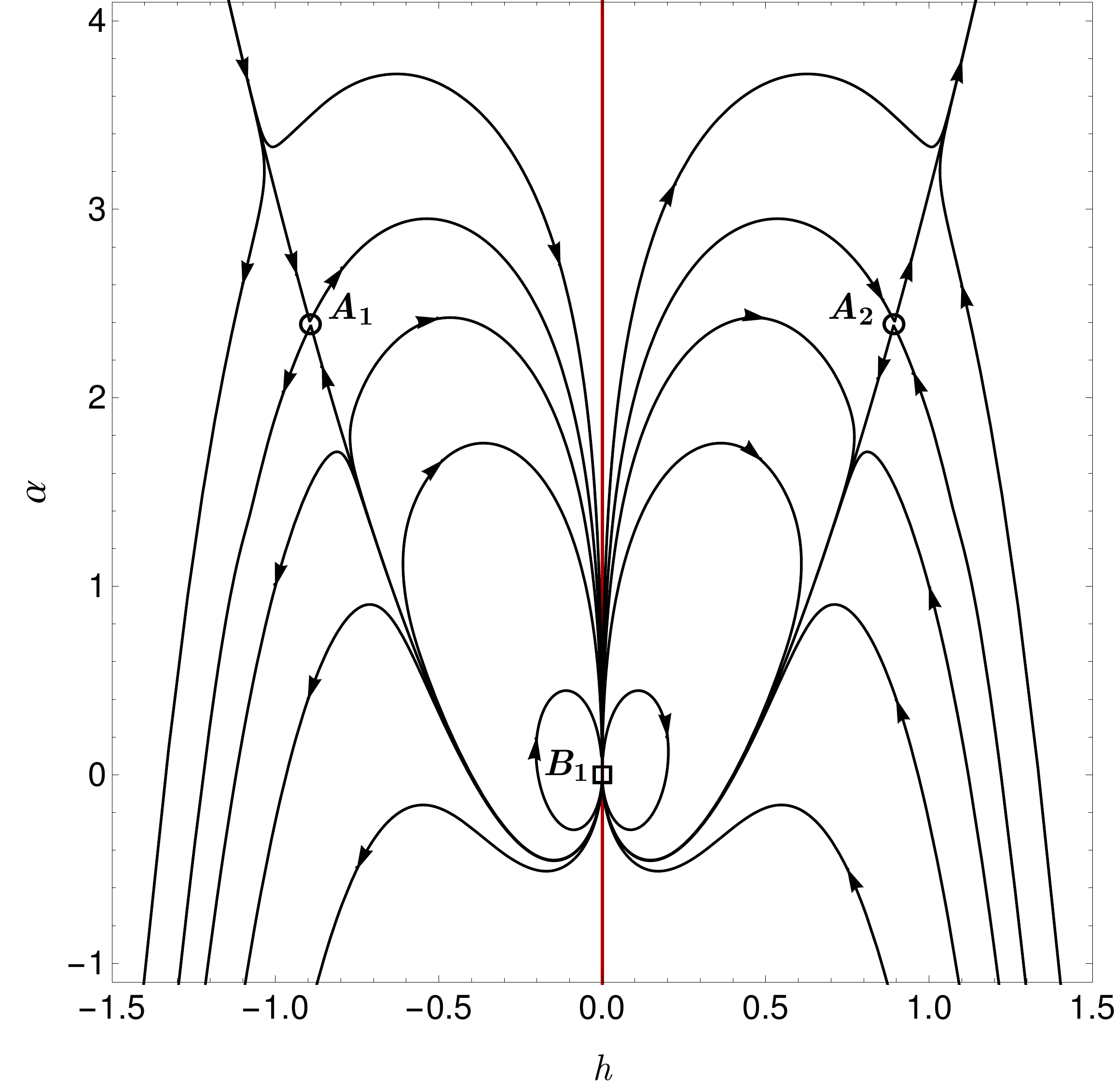}
\caption{\small The vector field \eqref{maineq} and its phase space diagram for
a flat empty universe with
exponential Lagrangian function \eqref{expf} and $\Omega_f=3/2$. The green and
red lines represent zero sets of the numerator and denominator of $v_1$
respectively. The blue parabola corresponds to $v_2=0$. Because of huge
variation, the vector lengths are not drawn to scale to better show the
discontinuity of direction at the singular line.}
\label{fig2}
\end{figure}

A very basic example illustrating these features is to take a flat, empty
universe and assume the exponential function
\begin{equation}
    F(\xi) = \Omega_f \mathrm{e}^{\frac{\xi}{\Omega_f}} - 1
    =\xi + \frac{\xi^2}{2\Omega_f}+\mathscr{O}(\xi^3),
    \label{expf}
\end{equation}
which includes the linear action, but no cosmological constant in the usual
sense. The specific form of $v_1$ is then
\begin{equation}
v_1\left(3\Omega_f\alpha,\sqrt{\tfrac12\Omega_f}h,a\right) =
    \frac{4\mathrm{e}^{-\alpha}+
    \mathrm{e}^{2\alpha}(3\alpha-3h^2-1)+
    3\mathrm{e}^{h^2}(h^4+(1-2\alpha)h^2+\alpha-1)}
    {\sqrt{2\Omega_f^3}(3\mathrm{e}^{2\alpha}+\mathrm{e}^{h^2})h},
\end{equation}
where the additional factors in the arguments are only introduced to shorten
the formula. It is still essentially transcendental, so one has to resort to
qualitative analysis first to locate the points and regions of interest. This
can be done with the help of Figure \ref{fig2}, which shows the planar
vector field $(v_1,v_2)$ together with the
locations of singular lines and zeros of the right-hand side $v$ (left panel),
and the phase portrait constructed from typical trajectories (right panel);
the particular value of $\Omega_f=3/2$ was chosen.

The left and right saddle points $A_1$ and $A_2$ correspond to time-reversed de Sitter
and standard de Sitter solutions, respectively, and their positions $(h_0,3h_0^2)$ are
given by $h_0 = \pm\sqrt{2\Omega_fw/3}$, where $w$ is the positive solution of
$\mathrm{e}^{-2w}+w=1$.

The singular critical point $B_1$ could
be considered as a static solution because it lies on the singular line $h=0$,
but also on the $W=0$ line, so in fact equation \eqref{modfried} is satisfied.
For the vector field, on the other hand, the limit at $B_1$ is not well
defined, as it depends on the path.

Importantly, there are no periodic orbits on either side of $B_1$, as the line
$h=0$ separates the neighbourhood of $B_1$ into two elliptic sectors of
opening $\pi$. The ``closed'' trajectories have $B_1$ as their limit point, so
they are asymptotically static both in the past and in the future.

More physically realistic evolutions here seem to consist of trajectories that are
attracted by $A_2$ and subsequently scattered along the unstable direction
towards infinity. These are expanding universes with ever increasing
acceleration, and also with initial singularity, which can be
read from the phase portrait: going back back in time, the trajectory has
increasingly negative $\alpha$, and discarding the exponentially small terms 
for large $h$ and $\alpha$ the right-hand side is approximately
\begin{equation}
    \dt{\alpha} = 12 h^3, \dt{h} = -h^2,
\end{equation}
making $\alpha$ and $h$ diverge in finite (negative) time.

\begin{figure}[h]
\includegraphics[width=.5\textwidth]{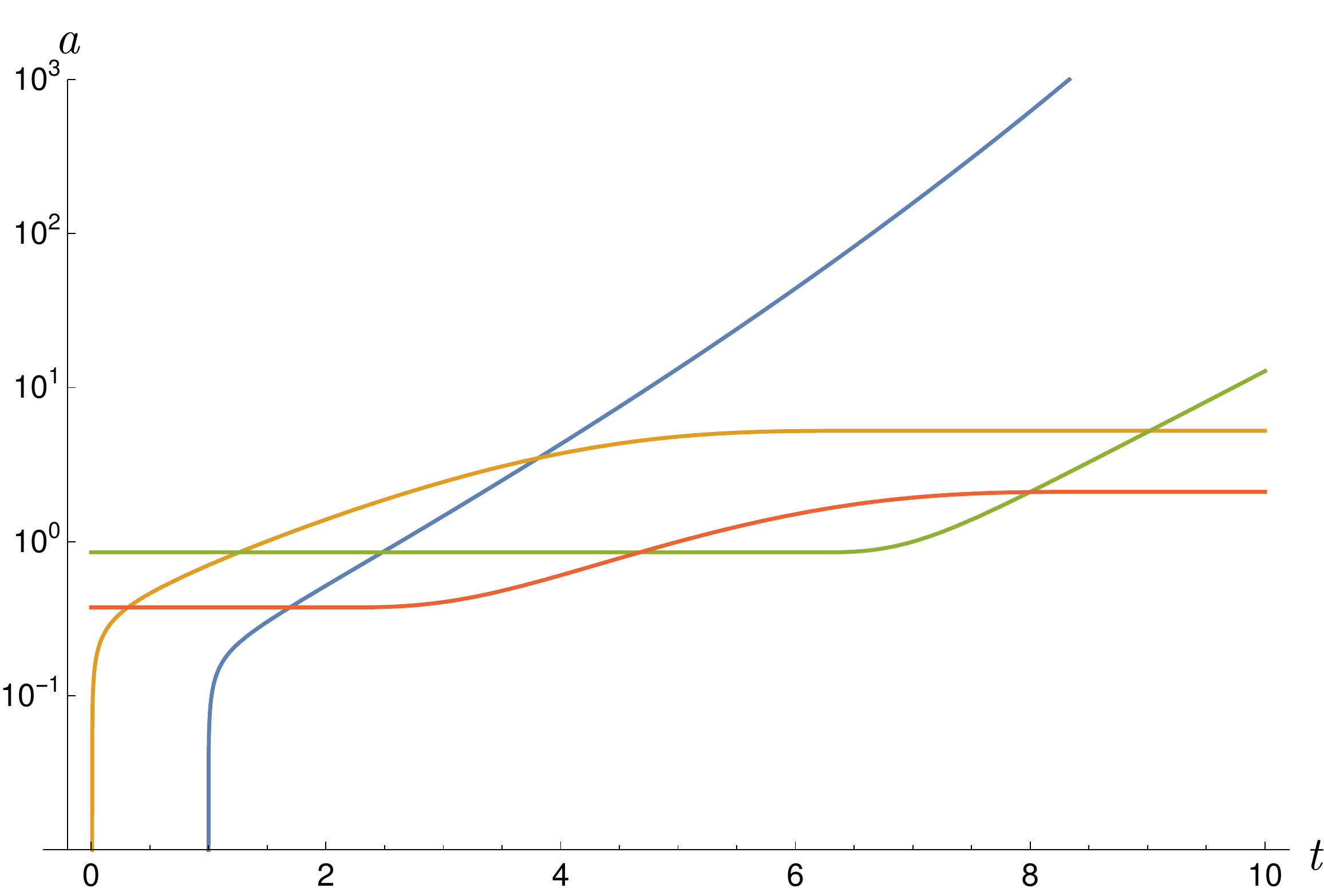}
\includegraphics[width=.5\textwidth]{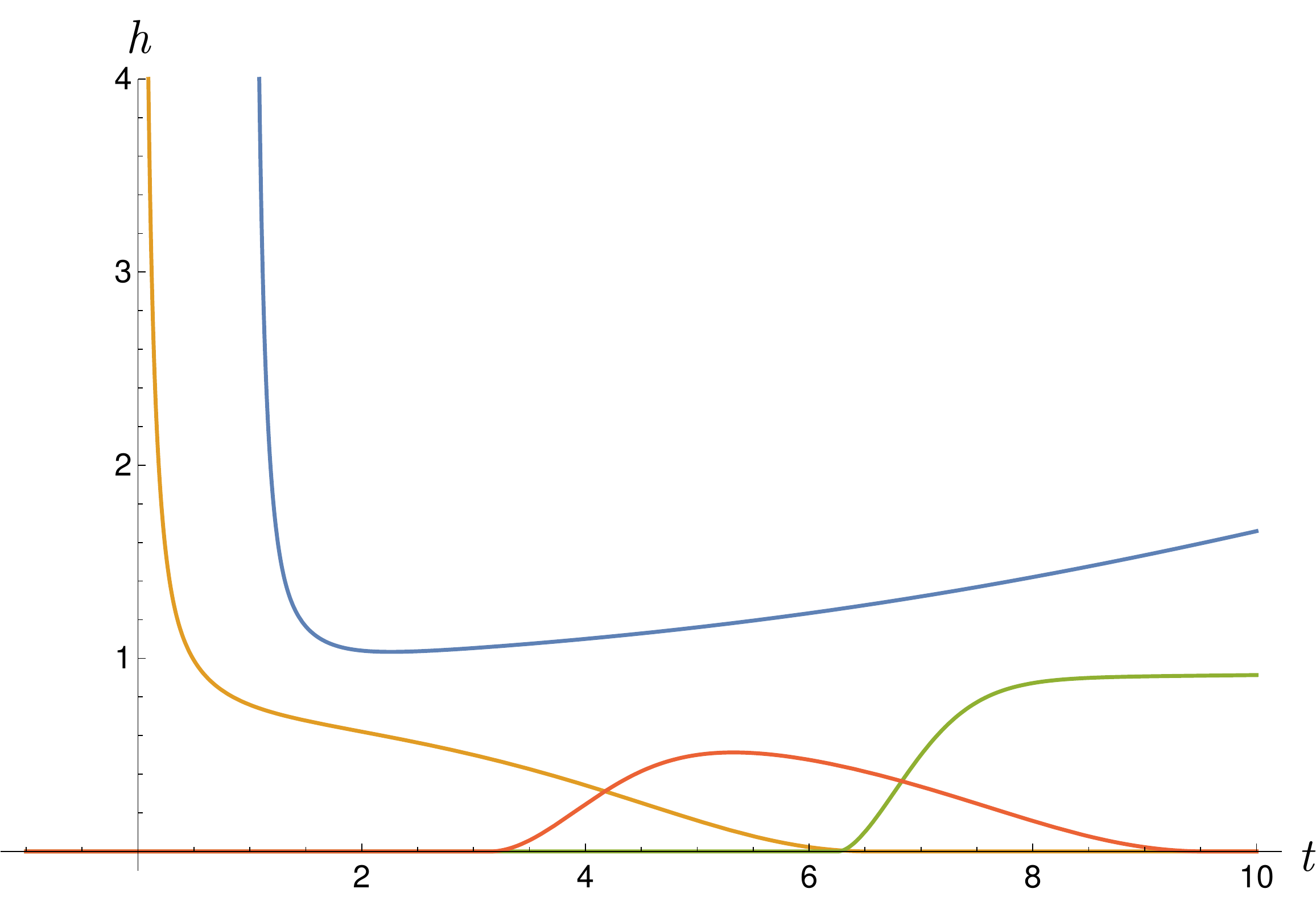}
\caption{\small Behaviour of $a$ and $h$ for typical 
flat, empty universes corresponding to \eqref{expf}. 
Blue and orange curves are trajectories which have infinite $h$ in the
past, but the former escapes to infinite $h$ while the latter is trapped by
$B_1$. Green and red curves both start at $B_1$ in the infinite past, but the former
escapes while the latter is recaptured.}
\label{fig5}
\end{figure}

There are also two mixed cases -- i.e., trajectories going from a big bang
becoming asymptotically static as they tend to $B_1$ and vice versa:
asymptotically static in the past, but then getting scattered by $A_2$ into
accelerated expansion. These exemplary behaviours of the scale factor and the
Hubble constant are plotted in Figure \ref{fig5}. Note that the time integration
constant $\tau_0$ such that $a(\tau_0)=1$ cannot always be chosen to make
$h(\tau_0)=1$, so it is adjusted for each trajectory for better visibility in this and
subsequent graphs.

It is probably more instructive to consider a more intricate model, which is
furnished by taking a rational function
\begin{equation}
    \tilde{f} = -1+\frac{\xi}{1-\xi^2} \quad\Longrightarrow\quad
    F(\xi) = -\frac{3\Omega_{\Lambda}}{2} + \xi + 
    \frac{4\xi^3}{9\Omega_{\Lambda}^2}+\mathscr{O}(\xi^5),
    \label{rationalf}
\end{equation}
which includes the constant term, so it can be identified with the
cosmological constant as in \eqref{oneparamF}. 
Note that if the series were to be used, different expansions in different
regions would be required. The reduction of the resulting powers of $R$ with
the characteristic polynomial would have to be carried out separately, which
would lead to cumbersome expressions -- if it were possible to obtain closed
ones at all.

\begin{figure}[h]
\includegraphics[width=.5\textwidth]{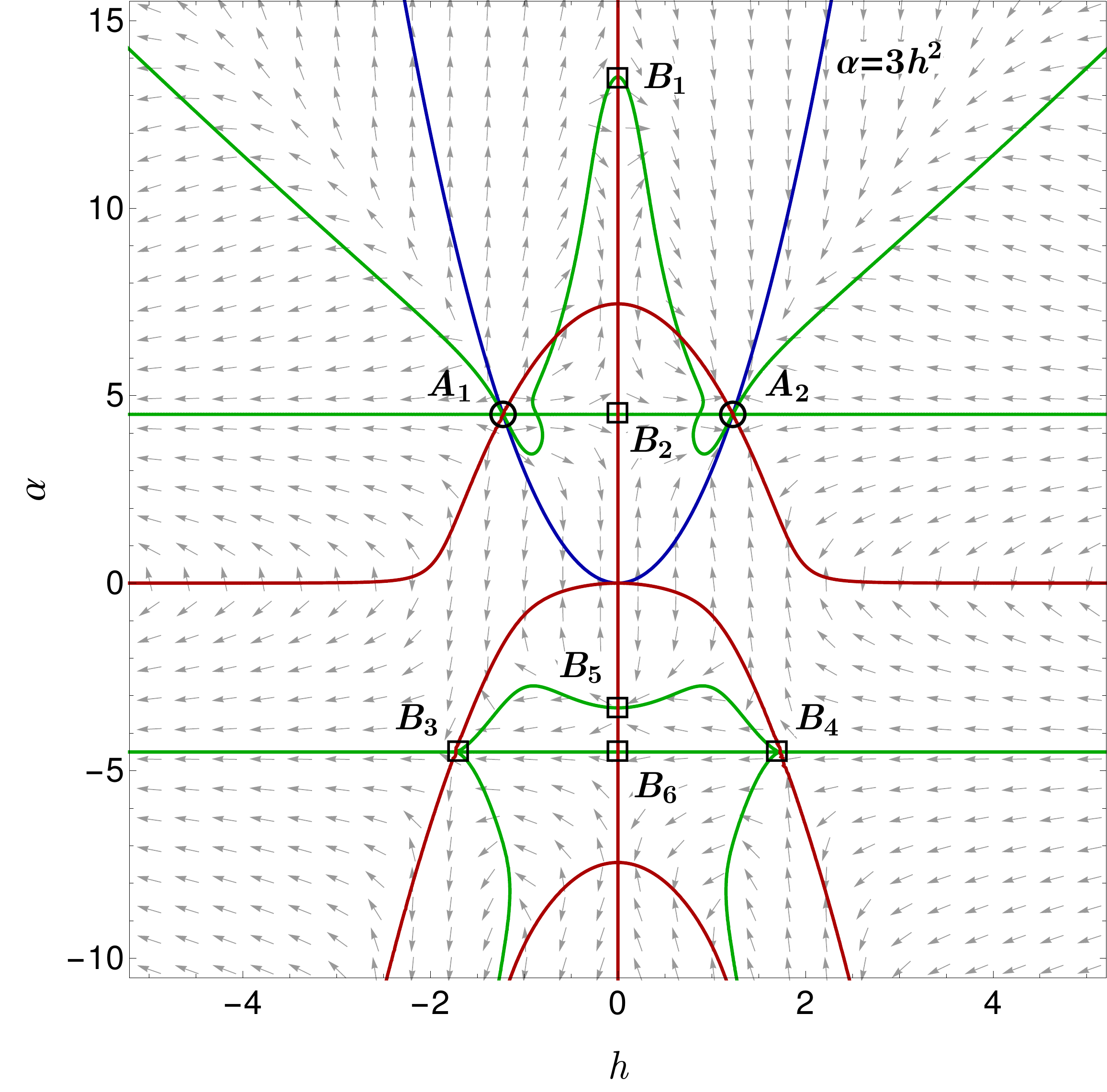}
\includegraphics[width=.5\textwidth]{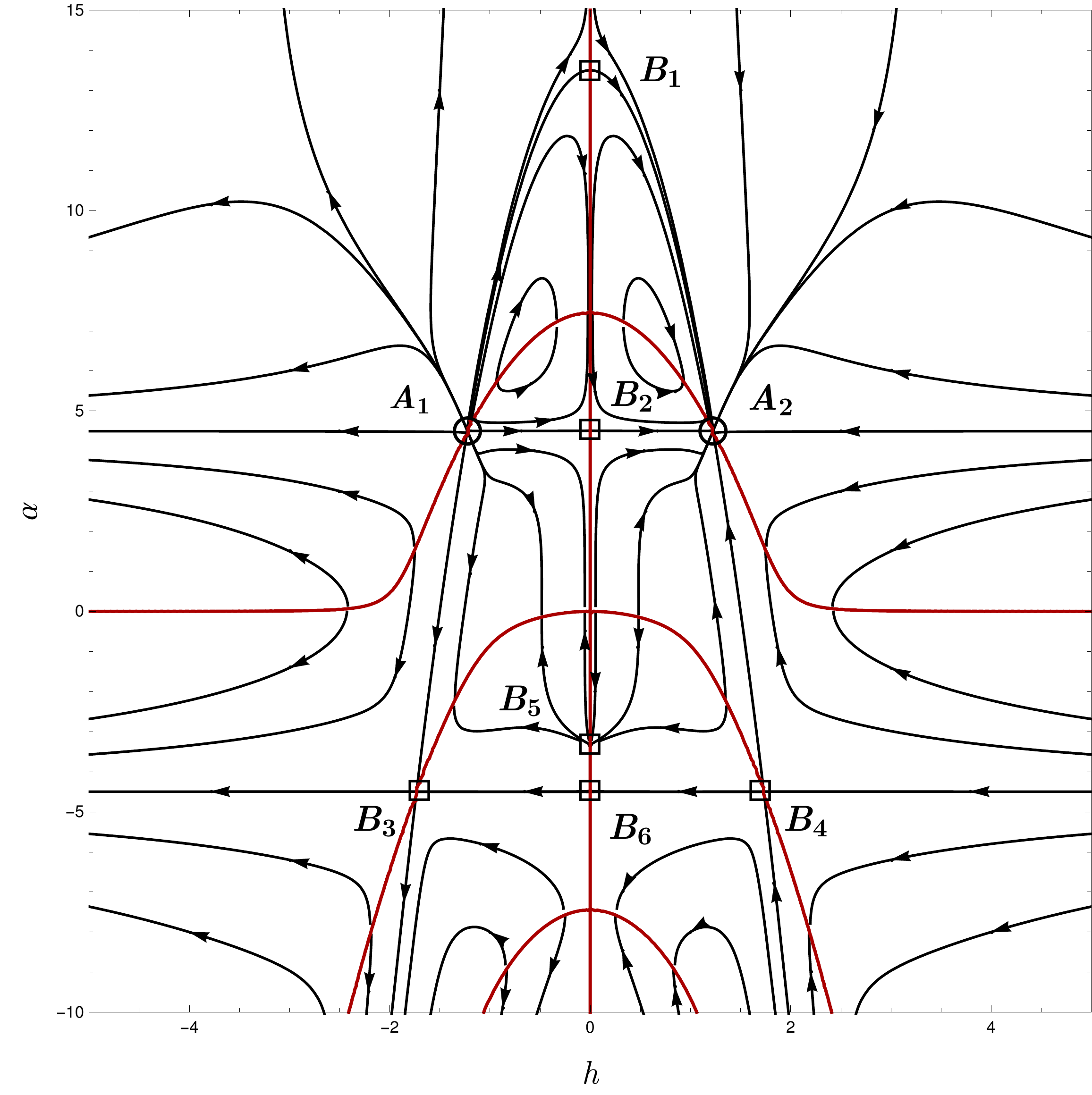}
\caption{\small The vector field \eqref{maineq} and its phase portrait for a
flat empty universe with
rational $F$ given by \eqref{rationalf} and with $\Omega_{\Lambda}=3$.
The field is singular on the red curves. The green curves represent a vanishing
numerator of $v_1$, and the blue parabola corresponds to $\protect\dt{h}=0$.
Because of huge variation, the vector lengths are not drawn to scale to better
show the discontinuity of direction at the singular lines.}
\label{fig3}
\end{figure}

Direct substitution of this $F$ into \eqref{maineq} produces a $v_1$ which is
several lines long, so it is perhaps best to skip its specific form and,
similarly to before, view the vector field and the various singular lines
of the phase space; they are shown in the left panel of Figure \ref{fig3}. The
picture is now
considerably more complex, with many more singular points of type $B$, for which
both the numerator and denominator in $\dt{\alpha}$ vanish. These points
signify a possible crossings through the otherwise impassable barriers
indicated by the red lines.

There are still only two critical points $A_1$, $A_2$ located at
$\left(\mp\sqrt{\tfrac12\Omega_{\Lambda}},\tfrac32\Omega_{\Lambda}\right)$,
which are asymptotic equilibria, and as before, they correspond to
time-reversed de Sitter and standard de Sitter solutions,
respectively. However, as the phase diagram of Figure \ref{fig3} shows, there
are now two heteroclinic trajectories connecting them, one through $B_1$ at
$\left(0,\tfrac92\Omega_{\Lambda}\right)$ and the other through $B_2$ at
$\left(0,\tfrac32\Omega_{\Lambda}\right)$.

There is a complication here, not present in the previous example, though. The
horizontal green lines at $\pm\tfrac32\Omega_{\Lambda}$ are singularities of
$F$, and
so also of the Friedmann equation, but they cancel out in $v_1$, resulting in
the straight-line trajectories. These are not singularities of curvature
either, because $\alpha$ and $h$ remain finite, so if one considers the action
principle as purely formal to obtain the dynamical equations, these solutions
could have some physical meaning. 

\begin{figure}[!ht]
\includegraphics[width=.5\textwidth]{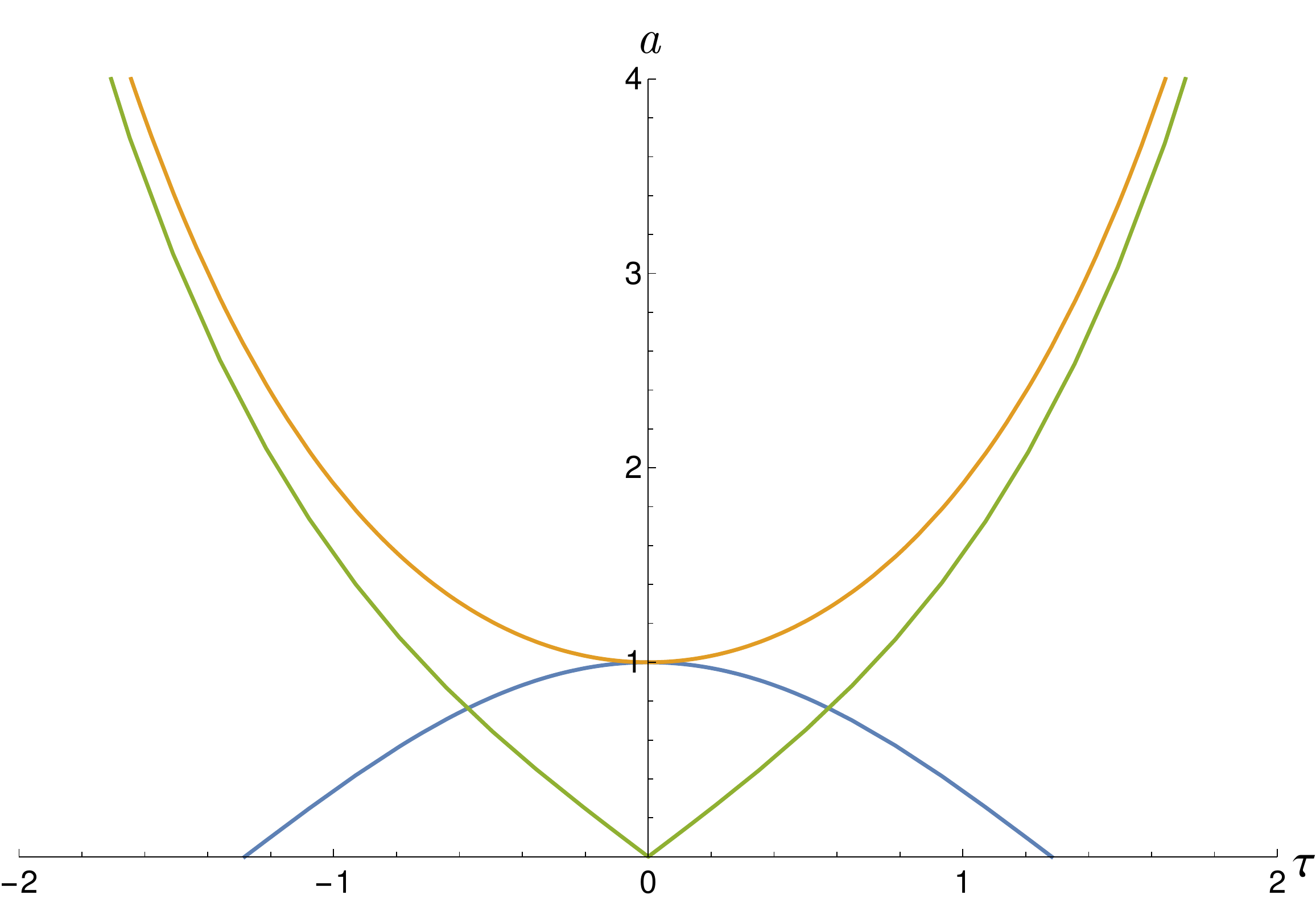}
\includegraphics[width=.5\textwidth]{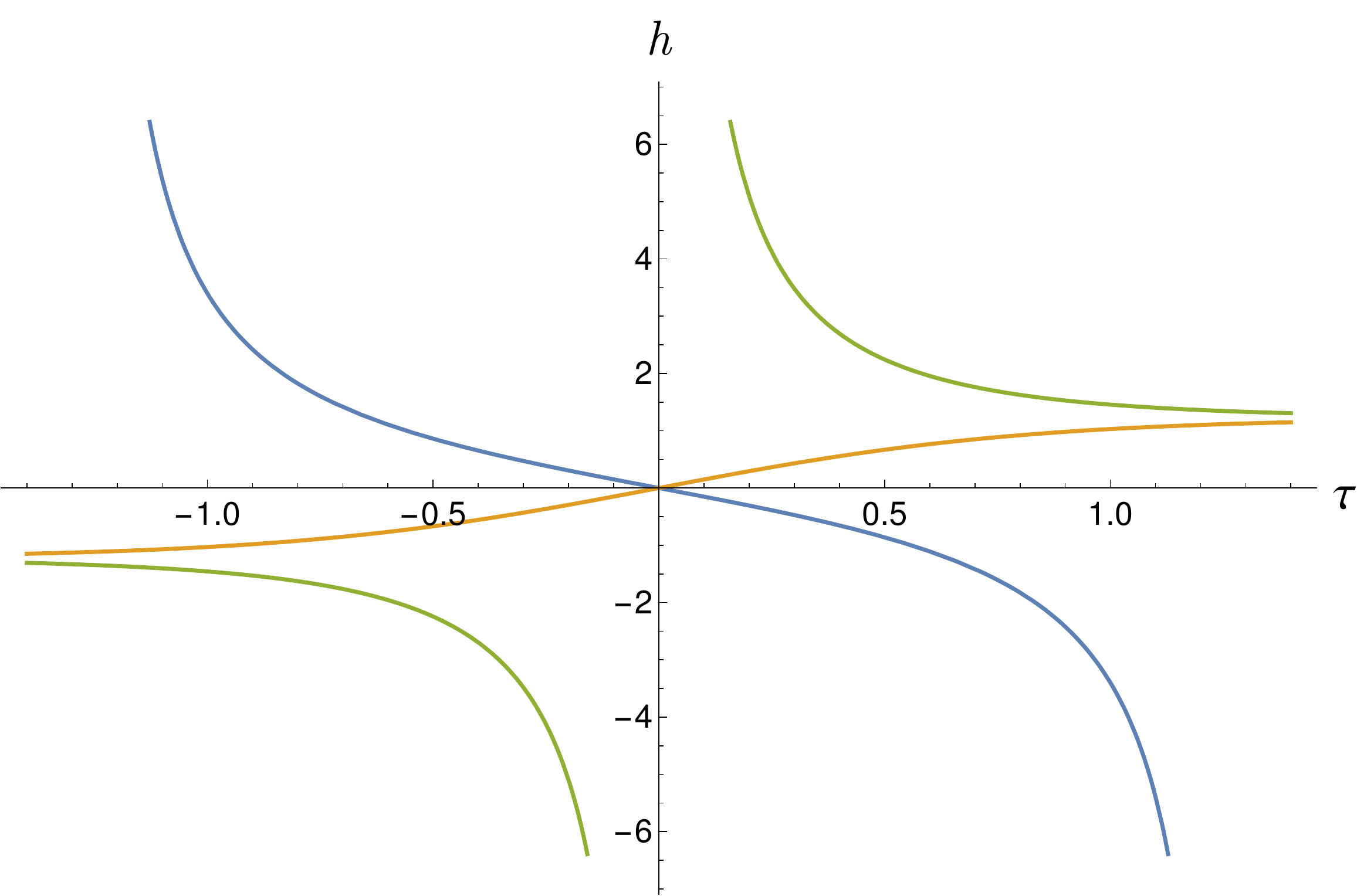}
\caption{\small Behaviour of $a$ and $h$ for the singular lines
$\alpha=\pm\tfrac32\Omega_{\Lambda}$ of Figure \ref{fig3}. The blue trajectory
goes through $B_4B_6B_3$, the orange one connects $A_1$ to $A_2$, and the green
one has $A_1$ and $A_2$ as limit points.}
\label{fig7}
\end{figure}

A similar situation is found for the pair $B_3$ and $B_4$ located at 
$\left(\pm\sqrt{\Omega_{\Lambda}},-\tfrac32\Omega_{\Lambda}\right)$, except
that the whole line can be
thought of as just one trajectory for which $h$ goes from $\infty$ to $-\infty$
in finite time. On both lines, the second equation $\dt{h}=v_2$ can be
integrated to give
\begin{equation}
    h = \sqrt{\pm\tfrac12\Omega_{\Lambda}}
    \tanh\left(\sqrt{\pm\tfrac12\Omega_{\Lambda}}(\tau-\tau_0)\right)
\quad\Longrightarrow\quad
    a = \cosh\left(\sqrt{\pm\tfrac12\Omega_{\Lambda}}(\tau-\tau_0)\right)
\label{explicit1}
\end{equation}
where the integration constant $\tau_0$ can be complex, giving in effect three
types of functions: tangent for the trajectory on the lower line, hyperbolic
tangent for
the $A_1A_2$ segment, and hyperbolic cotangent for the trajectories on the upper
line that escape to $\pm\infty$. The dependence of the scale factor and $h$ on
time for these cases is shown in Figure \ref{fig7}. Additionally, the
trajectories coming from infinity qualitatively reflect the behaviour of the
generic trajectories in the respective region in Figure \ref{fig3}; in
particular, the past singularity is reached in finite time.

\begin{figure}[!ht]
\includegraphics[width=.5\textwidth]{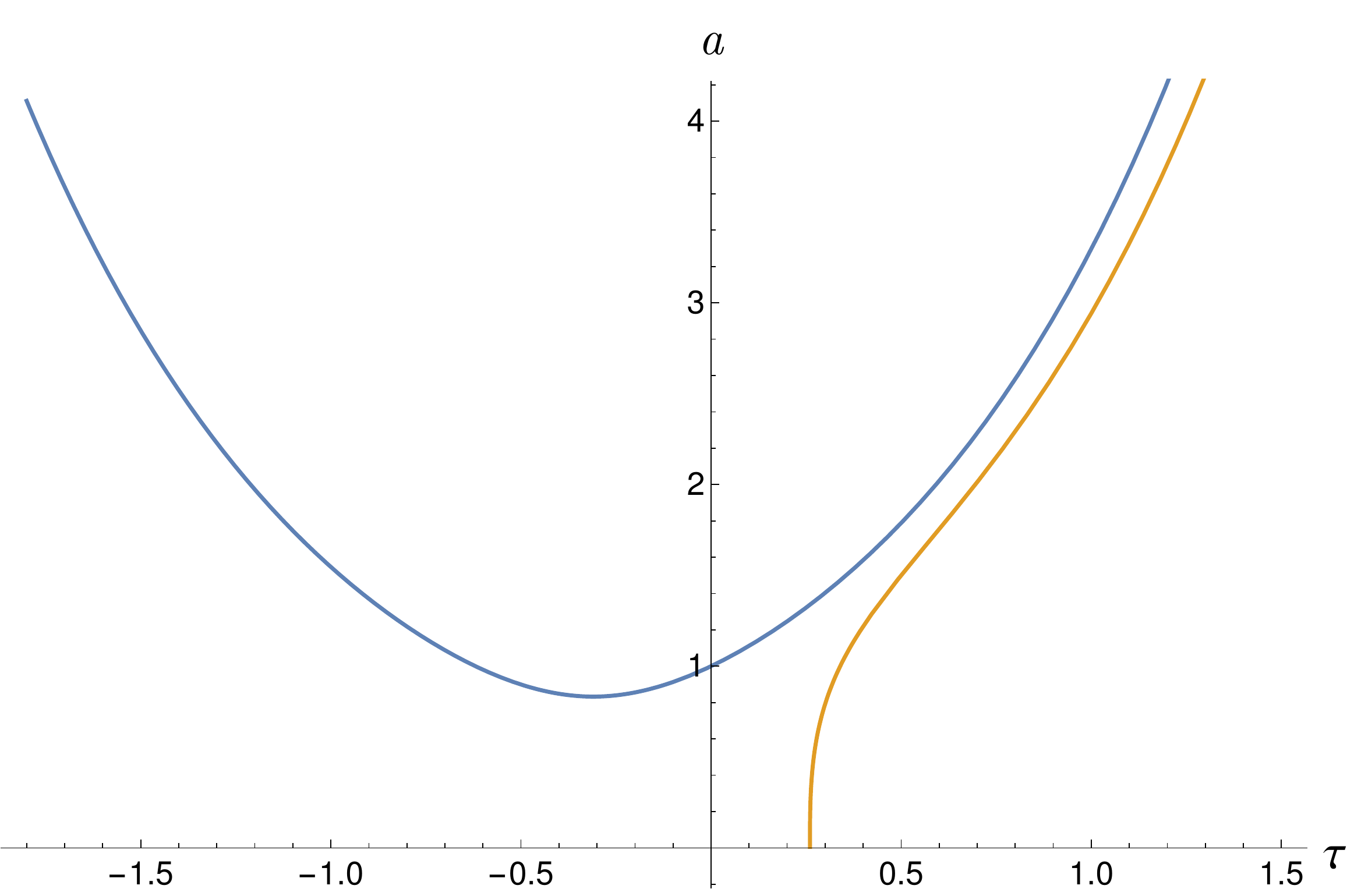}
\includegraphics[width=.5\textwidth]{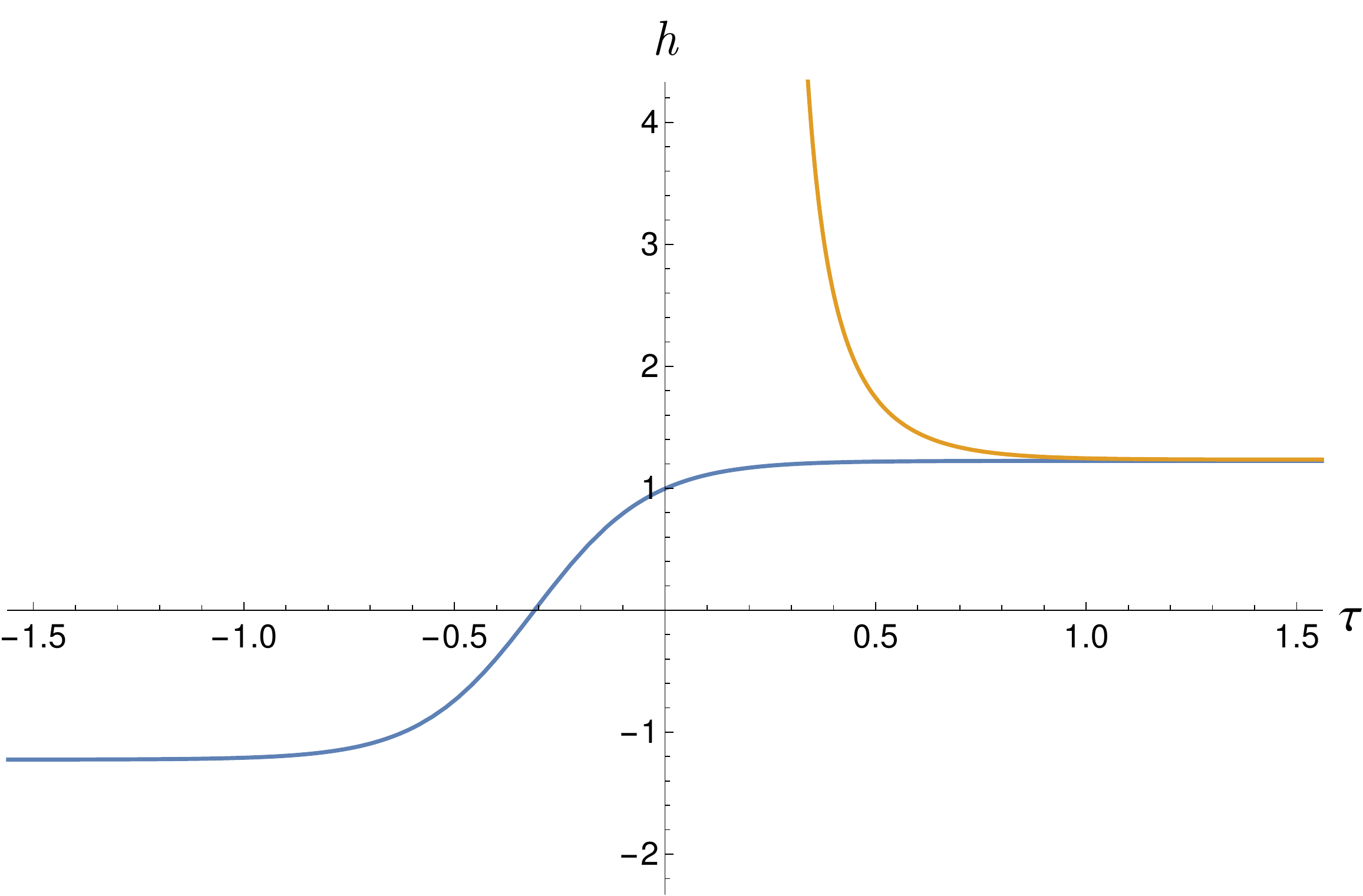}
\caption{\small The nontrivial heteroclinic trajectories of Figure \ref{fig3}:
the blue line corresponds to the one connecting $A_1$ and $A_2$ through $B_1$,
and the orange line to
the one coming from infinity to $A_2$ through $B_4$.}
\label{fig8}
\end{figure}

Outside the singular lines, there are the two special heteroclinic orbits: from
$A_1$ through $B_1$ to $A_2$ and from $B_4$ to $A_2$. The first is possible,
because the equation can be regularized by considering $\alpha$ as a
function of $h$ so that $\alpha'(h) = v_1/v_2$, which leads to a local
expansion at $B_1$
\begin{equation}
    \alpha = \tfrac92\Omega_{\Lambda}-6h^2+\mathscr{O}\left(h^4\right).
\end{equation}
This trajectory is similar to the one through $B_2$ but avoids the problem of
singular action. The second case, upon closer inspection, also admits
continuation through $B_4$, as is revealed by switching again to $h_1 :=
h-\sqrt{\Omega_{\Lambda}}$ as the
independent variable. The series for $\alpha$ can then be found
\begin{equation}
    \alpha = -\tfrac32\Omega_{\Lambda} - 12\sqrt{3}h_1
    +\mathscr{O}\left(h_1^2\right).
\end{equation}
Both of these solutions are shown in Figure \ref{fig8}, the first is probably
the best candidate for a ``bounce'' universe, and the second has a big-bang
singularity.

Looking more closely at the behaviour at infinity also reveals an
asymptotic relation of the form $\alpha\sim -6h^2$, which, together with the
two previous expansions, suggests looking for the equation of the extended
separatrix involving $\alpha+6h^2$.
Indeed, it turns out that there is a parabola through $B_3$, $A_1$, $B_1$, $A_2$ 
$B_4$ given by
\begin{equation}
    U: = \alpha+6h^2-\frac92\Omega_{\Lambda}=0,
    \label{inv_para}
\end{equation}
which is an invariant set, i.e,
\begin{equation}
    \left.\frac{\text{d}U}{\text{d}\tau}\right|_{U=0} = 0,
\end{equation}
as can be checked by direct substitution.

Eliminating $\alpha$ from $U=0$ leaves a simple Riccati equation 
$2\dt{h}=3\Omega_{\Lambda}-6h^2$, which
again gives trigonometric solutions for $h$ and $a$ akin to \eqref{explicit1}
-- in particular, for the big-bang type
\begin{equation}
    a = \sinh\left(\sqrt{\tfrac92\Omega_{\Lambda}}(\tau-\tau_0)\right)^{1/3}
     \;\sim\; (\tau-\tau_0)^{1/3},
    \label{invBB}
\end{equation}
which is the behaviour of the standard Friedmann cosmology with the so-called
stiff matter characterized by $p=\rho$. The same equation of state holds
also for a minimally coupled massless scalar field $\phi$ for which the energy
density is just the kinetic term $\rho=\frac12\dot{\phi}^2$, or approximately
when the potential term
can be neglected: $\frac12\dot{\phi}^2\gg V(\phi)$. This suggests a
correspondence analogous to that of standard $R^2$ theories, which are
conformally equivalent to scalar field cosmologies \cite{Whitt}.

The introduction of matter through a nonzero $\Omega$ term means that the system
\eqref{maineq} can no longer be simply visualized on a plane, but particular solutions can
still easily be obtained numerically. The most important
ingredient would be dust matter ($\gamma=1$), and following that, radiation
($\gamma=\tfrac43$), but since the latter constitutes a tiny fraction of
$\Omega$ in the standard $\Lambda$CDM model, $\Omega=\Omega_m a^{-3}$ was
assumed in the numerical integration. Thus, this particular model will depart
from reality close to the Big Bang by ignoring the radiationr-dominated GUT era and
the inflationary phase, when the value of $\Lambda$ is much larger
than the $\Lambda$CDM one used below.

\begin{figure}[h!]
{\centering
\includegraphics[width=0.6\textwidth]{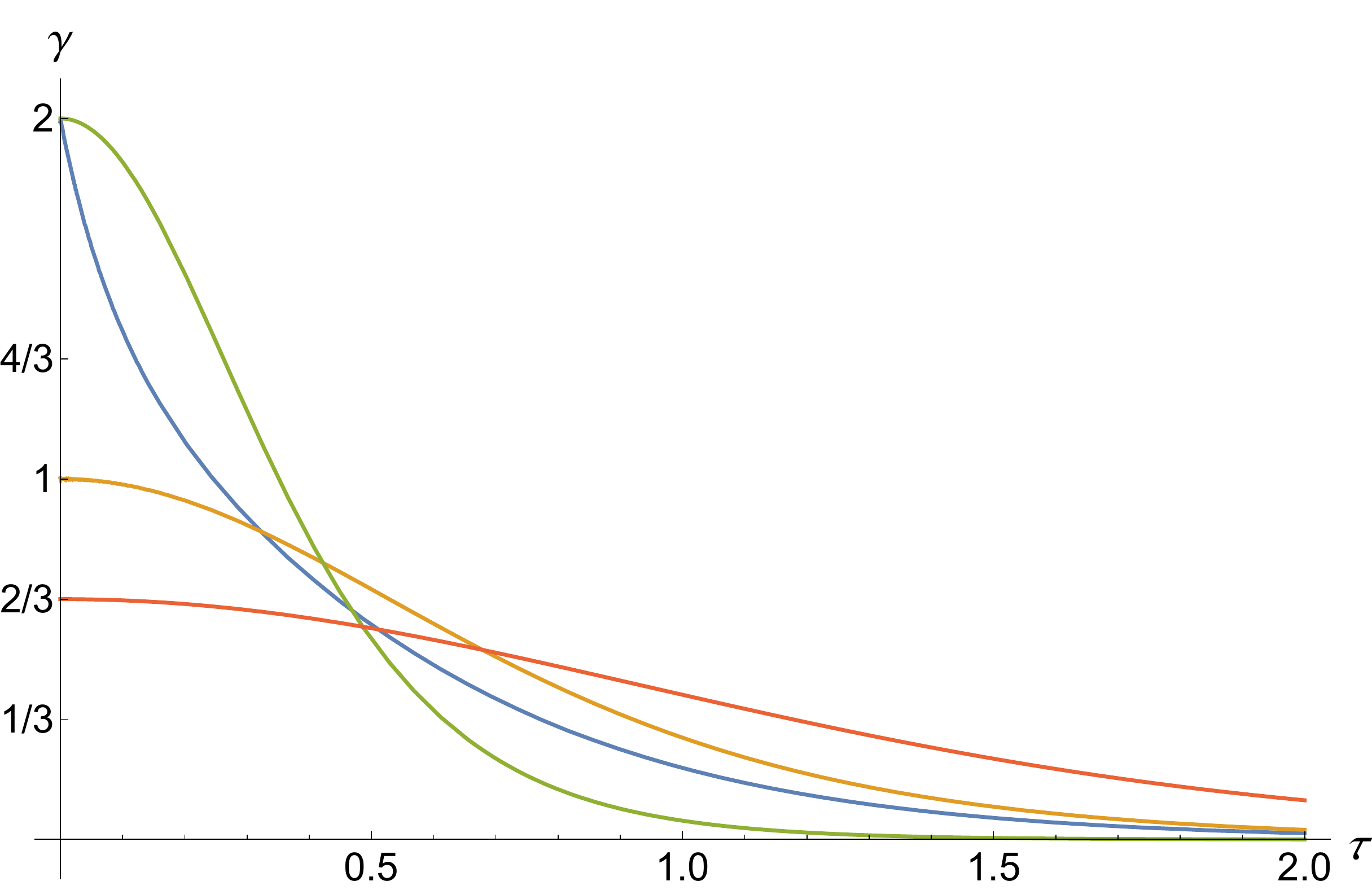}
\caption{\small The time-dependent index of the equation of state
$p=(\gamma-1)\rho$. The orange line corresponds to the $\Lambda$CDM model
\eqref{stiff_Fred} with $\Omega_{\Lambda}=0.69$ and $\Omega_m=0.31$; the blue
line is almost the same scenario, but with $\Omega_m=0.3$ and the addition of
$\Omega_s=0.01$. The green and red lines are big-bang solutions for a flat
universe with dust and rational $F$ given by \eqref{rationalf},
$\Omega_{\Lambda}=1.4$ and $\Omega_m=0.6$. They both tend to $A_2$: the former
through $B_4$ and the latter from the right of the first quadrant (see Figure
\ref{fig3}).}
\label{fig9}}
\end{figure}

A surprising property to notice is that the parabola
\eqref{inv_para} is still an invariant set, and accordingly equation
\eqref{invBB} gives a
bing-bang solution also with dust. This is due to the singular nature of the
denominator in the $\Omega/F''$ term in $v_1$. Although this means that the
stiff matter component
dominates in the earliest epochs, the ``effective equation of state''
$p/\rho$ changes with time as the de Sitter state is reached. By analogy with
the standard Friedmann cosmology, one can eliminate $p$ from the second
Einstein equation to obtain the time-dependent adiabatic index as
\begin{equation}
    \gamma_{\tau} = \frac23\left(1-\frac{a \ddt{a}}{\dt{a}^2}\right)
    = \frac23\left(1-\frac{\alpha}{3h^2}\right). 
\end{equation}
This function can be used to compare the behaviour of the density for the
present model and the corresponding Friedmann equation including the stiff
matter term, i.e.,
\begin{equation}
    h^2 = \Omega_{\Lambda} + \Omega_m a^{-3} + \Omega_s a^{-6},\qquad
    \Omega_{\Lambda} +\Omega_m + \Omega_s = 1.
    \label{stiff_Fred}
\end{equation}
The comparison is shown in Figure \ref{fig9}.

In the present case, there is no constraint on the sum of all the
$\Omega$ terms, and $\Omega_{\Lambda}$ and $\Omega_m$ need not be the same as
in the $\Lambda$CDM model, because the Einstein equations are different.
The parameter values are both subject to estimation from
observations, but for the present qualitative comparison one can use the
asymptotic behaviour of \eqref{invBB}:
$a\sim\exp(\sqrt{\Omega_{\Lambda}/2}\tau)$, which should correspond to the
relevant asymptotics of $\Lambda$CDM, i.e., $a\sim\exp(\sqrt{0.7}\tau)$, so
that $\Omega_{\Lambda}=1.4$ was chosen for the $f(R)$ equations.

At any rate, the comparison shows that the universe whose trajectory lies in
the first quadrant (Figure \ref{fig3}) and tends to the de Sitter attractor
$A_2$ has $\gamma=2/3$ during the big bang (red in Figure \ref{fig9}), so it
corresponds to cosmic
strings \cite{Dabrowski,Dabrowski2}. This is peculiar, because it means that the
matter term ($a^{-3}$) must be cancelled close to the initial singularity, so
that only the $a^{-2}$ term matters instead. It happens due to the trajectory
approaching the horizontal singular line
of $\alpha=3\Omega_{\Lambda}/2$ so asymptotically the solution
\eqref{explicit1} holds and $a\sim (\tau-\tau_0)$. The transition from cosmic
strings directly to exponential expansion makes this class of trajectories
unlikely as physical models.

The heteroclinic trajectory (green in Figure \ref{fig9}) is unchanged by dust
with $\gamma\approx 2$ stationary at first, then decaying to the
``dark energy'' level. This decrease is faster than for the corresponding
$\Lambda$CDM with stiff matter (blue), but the agreement is much better than
in the previous scenario. The shape resembles more that of
the standard $\Lambda$CDM (orange) in that there is no cusp, although different
types of matter dominate initially. This in itself is not an obstacle, as it is
unlikely that classical GR and dustlike matter determine the initial
singularity anyway, and in the bouncing scenarios $a''(0)=0$, so that $\gamma$
could even tend to infinity.

An interesting analogy here is that the heteroclinic trajectory is
unchanged by the addition of dust, so that it can be thought of as defined
purely by the geometry and the function $f(R)$ -- quite as the
cosmological constant can be thought of as a geometric term rather than an
actual material component. In both cases, such content-independent gravity only
makes sense as a model for the late homogeneous universe, not at smaller scales like
black holes. Note also that this particular example \eqref{rationalf} was
deliberately chosen with a singularity so that it cannot be treated
perturbatively. By itself, it may not be a replacement for $\Lambda$CDM, but
its most prominent feature, the invariant manifold $U=0$, appears as a guidepost
in further generalizations, partly because it effectively reduces the fourth-order
Einstein equations to an analogue of the Friedmann equation, which is easily
solvable. One goal of future investigations will thus be to find models where
such invariant curves exist and are nontrivially perturbed by matter.

Coming back to the general dynamics, an undesirable global feature of dynamics
with a singular $F(\xi)$ is that the
phase space is cut into several regions by the red lines and the trajectories
cannot be continued through them even with local analysis because the vector
field's directions are opposite on each side. Nevertheless, $A_2$ is a steady
state attractor for almost the whole first quadrant, and there are two
heteroclinic scenarios without singularities.

\begin{figure}[h!]
{\centering
\includegraphics[width=0.75\textwidth]{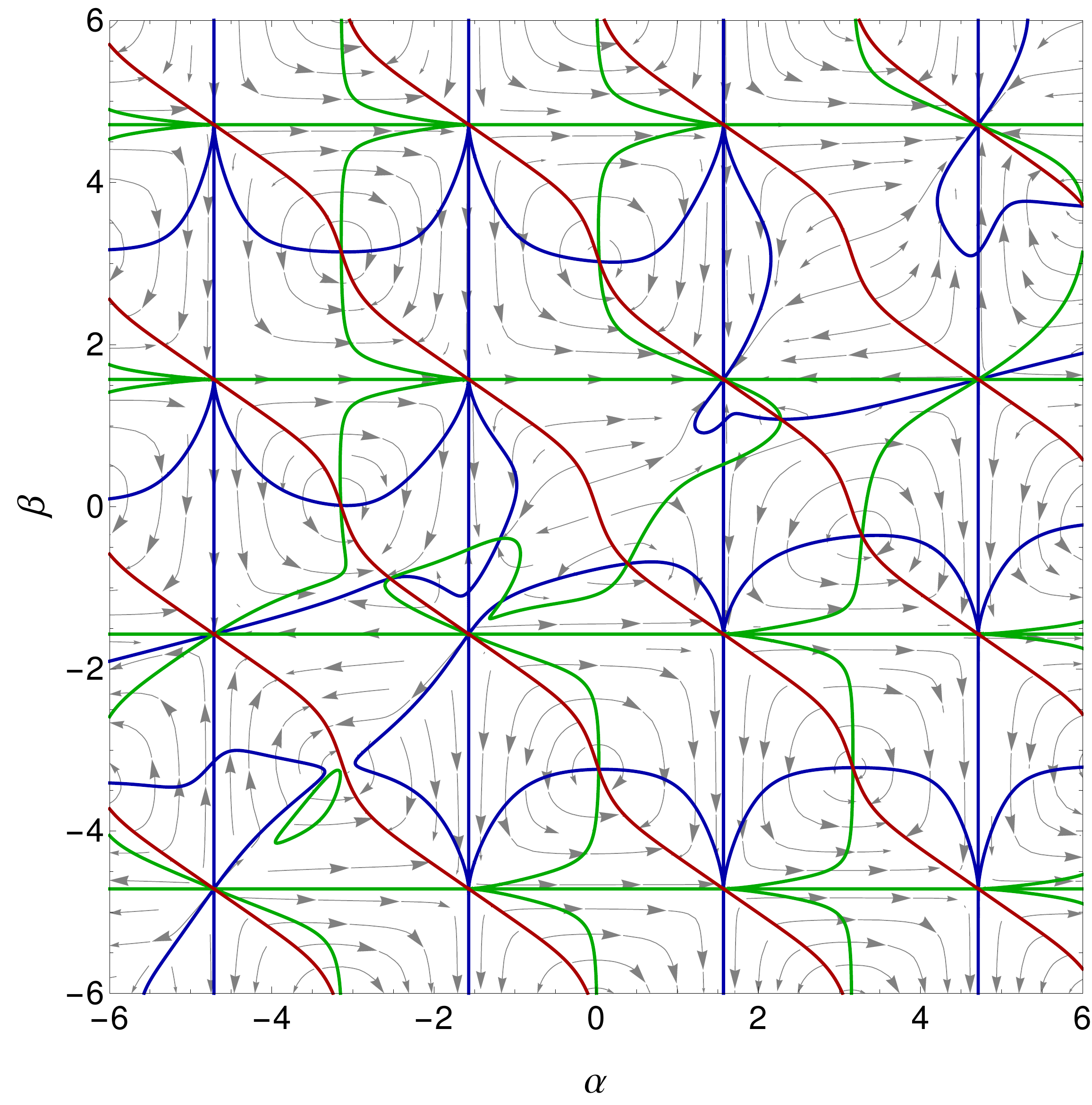}
\caption{\small The vector field and the singular lines for the system of
equations \eqref{trigsys} with trigonometric
$F(\xi)$ of \eqref{trigF} and $\Omega_{\Lambda}=1$.}
\label{fig6}}
\end{figure}

This behaviour is more pronounced when one considers more peculiar setups --
for example, with the periodic Lagrangian
\begin{equation}
    F(\xi) = \Omega_{\Lambda}\left( -\tfrac32+
    \tan\left(\frac{\xi}{\Omega_{\Lambda}}\right)\right).
\label{trigF}
\end{equation}
Because $F$ enters the equations with the rescaled eigenvalues $\alpha$ and
$\beta$ as its arguments, it is more convenient to eliminate $h$ and use the
eigenvalues as the dependent variables. In order to do that, a rescaled time
$\text{d}\sigma:=\text{d}\tau/h$ can be used, giving for the flat case
\begin{equation}
\left\{\begin{aligned}
   \frac{\text{d}\alpha}{\text{d}\sigma} &= 
    \frac{6\Omega+W}{3F''(\alpha)+F''(\beta)},\\ 
   \frac{\text{d}\beta}{\text{d}\sigma} &= 
    \frac{6\Omega+W}{9F''(\alpha)+3F''(\beta)}
    -(\alpha-\beta)\left(\tfrac13\alpha-\beta\right).
\end{aligned}\right.
\label{trigsys}
\end{equation}
This setup gives rise to a period cell structure of the phase space, as seen in
Figure \ref{fig6}, and there are infinitely many critical points and
heteroclinic orbits to choose from.

At present
this cannot be considered to be more than a toy model, but it hints at the
possibility of constructing a phase space with compartments for different
epochs of evolution separated by the singular lines and transitions taking
place through the critical points. The behaviours of $h$ and $a$ would need to
be recovered from that of $\alpha$ and $\beta$ in order to give physical
interpretation, and at first glance, it is hard to judge whether the complexity
comes from the choice of dependent variables, or is an intrinsic feature of the
tensor $f(R)$ theory.

The determination of the actual (real, if one can call it that) $F(\xi)$, or
$f$, is a
question in itself, and at present it is hard to imagine what other fundamental
theory could provide it. At the very least it should be constrained by
observations, but some new approach will be required not to merely fit
subsequent polynomial approximations of a series if one wants to recover the
complete function.

\section{Observational formulae}

In order to assess the applicability of the proposed construction one must turn
to observational cosmology. The detailed numerical analysis is outside the
scope of this article and will be deferred to future work. Nevertheless, some
preparatory analysis is straightforward and can be given here.

The standard cosmological test relies on the supernovae Ia data and the
relationship between the redshift and luminosity. In the Friedmann case, there
is a direct relation between $H^2$ and the redshift, so the integration of time
and distance is straightforward. Here, the equations involve
up to the third derivative of the scale factor, so another route needs to be
taken: for small redshifts, a series formula binding various expansion
coefficients can be given, while in the general case, the dynamical system has
to be integrated.

Recall first that the redshift is linked to the scale factor by
$z+1 = a^{-1}$, for $a(0)=1$ at present, and that the luminosity distance to an
object at comoving distance $r$ is $d_L = r (1+z)$. Provided, then, that $r$ can
be expressed by $z$, this will allow us to calculate the apparent luminosity and
relate to observations \cite{Weinberg}.

The required expression follows from the condition of the null geodesic:
$\text{d}s^2=0$, which for the metric \eqref{RW} gives directly
\begin{equation}
    r = \frac{1}{\sqrt{k}}\sin\left(\sqrt{k}\int\frac{\text{d}t}{a}\right),
\end{equation}
where a limit is understood for $k=0$. Assuming that $a$ or $z$ are monotonic
functions of $t$ that can be used for parametrization of the light path, the
above can be rewritten as
\begin{equation}
    d_L = \frac{1+z}{H_0\sqrt{\Omega_k}} \sin \left(\sqrt{\Omega_k}
\int_0^z\frac{\text{d}z}{h}\right).
\label{dL}
\end{equation}

In the standard model, $H$ is simply given as a function of $z$ by the
Friedmann equation, and the integral can even be explicitly calculated by means
of elliptic functions \cite{Dabrowski}. As mentioned above, this cannot be done
here, but following \cite{Dabrowski}, the main equation can be used to give
constraints of the higher characteristicss -- the deceleration parameter $q$ and
the jerk $j$:
\begin{equation}
    q := -\frac{\ddt{a}a}{\dt{a}^2} = -\frac{\alpha}{3h^2},\qquad
    j := \frac{\dddt{a}a^2}{\dt{a}^3} = \frac{\dt{\alpha}}{3h^3}-q.
\end{equation}

A change of the independent variable from $t$ (or $\tau$) to $z$ immediately
gives
\begin{equation}
    \frac{\text{d}h}{\text{d}z} = \frac{1+q}{1+z}h,\qquad
    \frac{\text{d}q}{\text{d}z} = \frac{j-2q^2-q}{1+z},
\end{equation}
which then allows us to expand $h$ in the integral \eqref{dL} in powers of $z$,
so that the whole expression can be expanded as 
\begin{equation}
    d_L = \frac{z}{H_0}\left( 1 +\frac{1-q_0}{2}z
        - \frac{1+j_0-3q_0^2-q_0+\Omega_k}{6}z^2+\mathscr{O}(z^3)\right).
\end{equation}
For small $z$, this provides a means to finding $H_0$, $q_0$ and $j_0$ from the
luminosity data, but one also has to take into account that these parameters
are not independent. In the standard model, $q$ can be eliminated because
$h'(z)$ is an explicit function of $z$ and the density parameters $\Omega$.
Similarly here, the jerk is constrained by the main equation, which for this
purpose becomes
\begin{equation}
    j + q = \frac{6\Omega+W(\alpha,\beta,h,a)}
    {3(3F''(\alpha)+F''(\beta))h^4},
\end{equation}
with
\begin{equation}
    \alpha = -3h^2q,\quad \beta = h^2(2-q) + \Omega_k(1+z)^2.
\end{equation}
So, given the function $F$, the constraint on $j_0$ is
\begin{equation}
    j_0 + q_0 = \frac{6\Omega_0 + W(-3q_0,2-q_0+\Omega_k,1,1)}
    {3(3F''(-3q_0) + F''(2-q_0+\Omega_k))}.
\end{equation}

Finally, to obtain the luminosity distance for larger redshifts, where a series
expansion is not practicable, an augmented dynamical system is a
straightforward solution. Assuming again that $z$ can be used as the
independent variable, as is the case in exponential expansion, a dynamical
equation for $d_L$ is necessary instead of the integral \eqref{dL}.

The null geodesic condition gives
\begin{equation}
    \frac{\text{d}r}{\text{d}z} = \frac{\sqrt{1-kr^2}}{H},
\end{equation}
and denoting the dimensionless distance by $l=H_0 d_L$ leads to
\begin{equation}
    \frac{\text{d}l}{\text{d}z} = 
\frac{l}{1+z}+\frac{\sqrt{(1+z)^2-\Omega_k l^2}}{h},
\end{equation}
while the basic system now reads
\begin{equation}
\left\{
\begin{aligned}
    \frac{\text{d}\alpha}{\text{d}z} &=
    -\frac{6\Omega+W\left(\alpha,\beta,h,(1+z)^{-1}\right)}
    {(1+z)(3F''(\alpha)+F''(\beta))h^2},\\
    \frac{\text{d}h}{\text{d}z} &= \frac{3h^2-\alpha}{3(1+z)h}.\\
\end{aligned}\right.
\end{equation}
Because $z$ has become the independent variable, this system is non-autonomous and
only two-dimensional (regardless of $k$ and $\Omega$). Even in the Friedmann
case, for more complex $H(z)$, the integral \eqref{dL} has to be obtained
numerically. The only complication here is that three ordinary differential
equations need to be integrated; their initial conditions follow from the
definitions
\begin{equation}
    l(0) = 0,\quad
    \alpha(0) = -3q_0,\quad
    h(0) = 1.
\end{equation}

\section{Conclusions}

The main modification of the gravitational action proposed here is to include
terms nonlinear in curvature, but going further than polynomials, so that
rational functions with a finite radius of convergence or even transcendental
functions can be used.
Additionally, instead of considering just a function of the Ricci scalar
$f(\text{tr}[R])$, the
whole tensor can be treated as an argument, and the trace taken at the very
end to produce a scalar Lagrangian density $\text{tr}[f(R)]$. In the case of
transcendental
functions, this considerably changes the results, when compared to the 
ordinary $f(\mathscr{R})$ theories.

With a view to fully general treatment, such as including spin, the presented
derivation is valid for affine connections with nonvanishing torsion and
without the assumption of metricity. An important consequence is that for
nonsymmetric Ricci tensors one can no longer introduce an obvious metric
conformal to the original $g_{ab}$. This stems from the nonlinear functions of the
Ricci tensor entering the equations, instead of just functions of the Ricci
scalar multiplying $R_{ab}$ or $g_{ab}$.

Despite the difficulties, workable equations can be derived and applied to
the Robertson-Walker geometry so that the analogue of the standard cosmological
model may be studied. As is generally the case, the modified Einstein equations
are of higher order, and instead of one Friedmann equation, one has a
three-dimensional dynamical system.

An obvious complication is that the dynamical variables enter the equations both
inside and outside the transcendental functions, which leaves little hope for
explicit solutions. Nevertheless, these models are within reach and if the function
$f$ is determined from other fundamental principles, the dynamics and
observational consequences can still be effectively analysed, as shown here.

The analysis of phase portraits for both rational and transcendental $f$
reveals critical points which are attractors and which correspond to 
de Sitter solutions. More importantly, there also exist non-singular
``big bounce'' evolutions, which are heteroclinic trajectories, and explicit
solutions for them can be given. For the dynamical systems to be
two-dimensional it was assumed that the curvature was zero and no
ordinary matter was present. On the one hand this allows for a complete
visualization of the phase diagram, but on the other, it limits the physical
applicability. Still, the late or present Universe with accelerated expansion
can be modelled as the de Sitter attractor, while for the big bounce solutions
the scale factor does not approach zero, so that matter density never dominates
and neglecting it is justifiable.

If dustlike matter is included, the separatrix of the above
simplified rational model survives and the same explicit solutions hold. One
still has both big bounce and big bang solutions, not unlike those of
$\Lambda$CDM with stiff matter. In general, matter changes the early evolution
around the separatrix but not on it. Thus, the next possible step in
constructing a viable model seems to be identifying $f(R)$ such that it also
has an invariant submanifold, but which depends on $\Omega_m$, not just on the
geometry and $\Lambda$.

In any case, the elegant feature here is that the cosmological constant can
appear naturally because of how the theory is constructed -- it is identified
with the constant term of $f(R)$. Yet, even when this term was zero
($f=\exp{}-1$), the same sort of accelerated expansion appeared.

A more detailed study of the initial singularity in the presence of matter and
curvature index $k$ could lead to more interesting results still.
For example, seeing how one of the scenarios imitates stiff matter, it will be
interesting to ask if such cosmologies can be equivalent to standard general
relativity with a scalar field, similarly to the ordinary $R^2$ case. It is also
the quadratic $f(R)$ case for the Robertson-Walker geometry, when there is an
equivalence with the ordinary $f(\mathscr{R})$, although it does not seem to
extend to higher orders. Another convergence is found when the traceless Ricci
tensor vanishes, so that $R_{ab}$ is proportional to $g_{ab}$ and the Einstein
equations for both theories coincide. However, as the examples show, even for
an empty universe this might correspond only to fixed points, not to general
solutions of the full theory.

With a view to future work, some observational formulae are also given, so that
the basic cosmological tests can be applied. A comparison to the standard model
is in order to help guide the subsequent theoretical developments.
Specifically some constraints on the function $f$ should be obtained. The
crudest way would be to fit the first coefficients of its expansion, but
of course there is no hope in recovering the whole series this way.

Rather, one might want to
approach the problem by trying to fit a differential equation satisfied by $f$.
Already for linear differential equations with rational coefficients this would
reduce the number of parameters to finite, while at the same time allowing for
a the vast family of (confluent) hypergeometric functions and their
generalizations.

Future investigations could also address the question of reduction of the order
of the dynamical system \eqref{maineq}. For the Einstein-Hilbert action, the
third derivative of the scale factor does not enter, and only the Friedmann
equation, which is a relation between $H$ and $a$, is left. Here, the equation
involving the third derivative of the scale factor, or $\dt{\alpha}$, would be
reduced if $3F''(\alpha)+F''(\beta)=0$. For independent $\alpha$ and $\beta$
this happens only if $F$ is linear, so that GR is recovered.

If, on the other hand, there is a relation $\beta=\psi(\alpha)$,  then a nontrivial
solution to the functional equation $3J(\alpha)+J(\psi(\alpha))=0$ could
potentially be found. Such a relation is in itself a second-order differential
equation for the scale factor, so the dynamics is simplified, but it then also
means that the function $F$ is determined by $F''(\xi)=J(\xi)$. 

Ideally however, the function $f$ should be mainly constrained by experiment
not just the simplicity of the resulting equations. If this theory passes the
basic cosmological tests, analysing it in a wider context of gravitational
physics will help address this issue. Questions of instabilities
will have to be answered, although as suggested by \cite{Olmo}, the Palatini
approach, applicable here, provides a setting to avoid at least the
Ostrogradski instability. In general, issues such as ghost fields,
semiclassical stability and post-Newtonian (Solar System) tests will be
required, and hopefully undertaken, to ascertain the overall viability of the
presented extension.

\end{document}